\documentclass[12pt]{iopart}
\usepackage{iopams}
\usepackage{epsfig}
\eqnobysec
\begin{document}
\title[Free probability meets supersymmetry]
{Zooming in on local level statistics by\\ supersymmetric extension
of free probability}
\author{S.\ Mandt and M.R.\ Zirnbauer}
\address{Institut f\"ur Theoretische Physik, Universit\"at zu
K\"oln, Z\"ulpicher Stra{\ss}e 77, 50937 K\"oln, Germany}
\begin{abstract}
  We consider unitary ensembles of Hermitian $N \times N$ matrices
  governed by a confining potential $NV$ with analytic and uniformly
  convex $V$. From work by Zinn-Justin, Collins, and Guionnet and Maida it
  is known that the large-$N$ limit of the characteristic function for
  a finite-rank Fourier variable $K$ is determined by the Voiculescu
  $R$-transform, a key object in free probability theory. Going beyond
  these results, we argue that the same holds true when the
  finite-rank operator $K$ has the form that is required by the
  Wegner-Efetov supersymmetry method. This insight leads to a potent new
  technique for the study of local statistics, e.g., level
  correlations. We illustrate the new technique by demonstrating
  universality in a random matrix model of stochastic scattering.
\end{abstract}
\pacs{02.30.Fn, 02.50.Cw, 03.65.Nk, 05.30.Ch}
\vspace{2pc} \noindent{\it Keywords}: Random matrices, supersymmetry
method, free probability, universality in stochastic scattering

%
\maketitle

\section{Introduction}

The notions of `free probability' and `freeness' of non-commutative
random variables were introduced by Voiculescu in the study of
certain algebras of bounded operators \cite{VDN}. The word freeness
in this context means a kind of statistical independence of
operators. The algebraic concept of freeness of random variables has
a natural realization by random matrices in the limit of infinite
matrix dimension \cite{voicul2}; this realization is what we study,
develop, and apply in the present paper.

A central tool of the free probability formalism is the so-called
$R$-transform, which resembles the logarithm of the characteristic
function for commutative random variables. Voiculescu \cite{voicul1}
defined it by the functional inverse of the average trace of the
resolvent operator. A second approach to the subject is due to
Speicher \cite{speicher}, who expressed the moments of the random
matrix directly in terms of the Taylor coefficients of the
$R$-transform. Speicher's concept of non-crossing partition is a
mathematical expression of the dominance of planar Feynman graphs
(using physics parlance) in the large-$N$ limit. In the present paper
we will encounter both approaches: the analytical one of Voiculescu,
and the combinatorial one of Speicher.

Our long-term goal is a comprehensive description of the spectral
correlation functions and, ultimately, a proof of the universality
hypothesis which is expected for certain random matrix ensembles in
the large-$N$ limit. Although the $R$-transform is a powerful tool to
tackle density-of-states questions, it is fair to say that free
probability theory has not yet contributed much to our understanding
of the universality of spectral correlation functions at the scale of
the level spacing.

Bearing this in mind, we now change subject and turn to the so-called
supersymmetry method, by which we mean the technique of integration
over commuting and anti-commuting variables pioneered by Wegner
\cite{wegner} and Efetov \cite{efetov}. In its original formulation
(using the Hubbard-Stratonovich transformation) this method was
limited to Gaussian disorder distributions. Nonetheless, with this
limitation it has enjoyed great success in producing non-trivial
results for a number of physics problems, e.g., level statistics of
small metallic grains, localization in thick disordered wires,
scaling exponents at the Anderson transition, etc. In the present
paper we will take advantage of a recent variant, called
superbosonization, which makes it possible in principle to treat a
class of disorder distributions much wider than the Gaussian one.

Since their inception in the 1980's, free probability theory and the
supersymmetry method have coexisted with little or no mutual
interaction. Forecast in a prescient remark by P.\ Zinn-Justin
\cite{pzj-PRE}, the message of the present paper is that a new
quality emerges when the two formalisms are combined. More
specifically, we will show that the characteristic function of the
probability law of the random matrix ensemble -- an object of central
importance to superbosonization -- has a large-$N$ limit which is
determined by the $R$-transform. This result paves the way for a
number of applications. As a first application, we will illustrate
the new method by demonstrating universality for a random matrix
model of stochastic scattering.

\subsection{Summary of results}

Our results make reference to the $R$-transform, which we now
introduce in more detail. Consider the average trace of the resolvent
operator, $g(z) := \lim_{N \to \infty} N^{-1} \langle \mathrm{Tr} \,
(z-H)^{-1} \rangle$, $z \in \mathbb{C} \setminus \mathbb{R}\,$, of
the random matrix $H$. Voiculescu inverts the function $z \mapsto
g(z)$ to define the $R$-transform as the regular part of $g(z) = k
\mapsto k^{-1} + R(k) = z\, $. It has a power series $R(k) = \sum_{n
\geq 1}\, c_n\, k^{n-1}$ whose coefficients $c_n$ are called free
cumulants. These are analogs (in the non-commutative setting) of the
usual cumulants in that they are linear with respect to free
convolution.

In the present paper, we consider $\mathrm{U}_N$-invariant
probability measures $\mu_N$ with a density of the form
\begin{equation}\label{eq:def-mu(H)}
    d\mu_N(H) := \mathrm{e}^{-N\,\mathrm{Tr}\,V(H)} d H \;,
\end{equation}
where $dH$ denotes Lebesgue measure on the linear space of Hermitian
$N \times N$ matrices, and $\mathbb{R} \ni x \mapsto V(x)$ is
analytic and uniformly convex. The focus of our analysis is the
characteristic function
\begin{equation}\label{eq:char-func}
    \Omega(K) = \int \mathrm{e}^{\mathrm{Tr}\, H K} d\mu_N(H) \;.
\end{equation}
Note that $\Omega(K)$ is invariant under conjugation $K \mapsto
g^{-1} K g$ by $g \in \mathrm{GL}_N \,$.

Motivated by the supersymmetry method as reviewed in Section
\ref{sect:motiv}, we make a study of $\Omega(NK)$ for linear
operators $K$ of the form
\begin{equation}\label{eq:def-K}
    K = \sum_{a=1}^p \varphi_a \otimes \widetilde{\varphi}_a
    + \sum_{b=1}^q \psi_b \otimes \widetilde{\psi}_b \;,
\end{equation}
with $p, q$ kept fixed in the limit $N \to \infty$. Here $\varphi_a\,
, \psi_b \in \mathbb{C}^N$ are vectors, while $\widetilde{\varphi}_a
\,, \widetilde{\psi}_b \in (\mathbb{C}^N)^\ast$ are co-vectors. The
components of $\widetilde{\varphi}_a$ are complex commuting variables
or generators of the symmetric algebra $\mathrm{S}(\mathbb{C}^N)$,
whereas the components of $\widetilde{\psi}_b$ are complex
anti-commuting variables or generators of the exterior algebra
$\wedge(\mathbb{C}^N)$. The same statement, albeit with
$\mathbb{C}^N$ replaced by $(\mathbb{C}^N)^\ast$, applies to
$\varphi_a$ and $\psi_b $ respectively. All inner products $\langle
\widetilde{\varphi}_a , \varphi_b \rangle$ for $a,b=1,\ldots,p$ are
kept fixed in the large-$N$ limit.

For such $K$ we argue that the following holds:
\begin{equation}\label{eq:result-MZ}
    \lim_{N \to \infty} N^{-1} \ln\, \Omega(N K) =
    \sum_{n=1}^\infty \frac{c_n}{n}\; \mathrm{Tr}\;K^n\;.
\end{equation}
We note that this formula makes perfect sense as long as the power
series of the $R$-transform has infinite radius of convergence. The
latter property is ensured by our assumptions on $V$.

\subsubsection{Related mathematical work.}

To put the result (\ref{eq:result-MZ}) into context, we now mention
related mathematical work on the large-$N$ asymptotics of the
following spherical integral (known in the physics literature as the
Itzykson-Zuber integral):
\begin{equation}\label{eq:HCIZ}
    I_{X_N}(k) = \int_{\mathrm{U}_N} \mathrm{e}^{k \,
    \mathrm{Tr}\, (X_N\, g \Pi g^{-1})} dg \;,
\end{equation}
for the case of a rank-one projector $\Pi$. Let the eigenvalues
$x_{1,N}, \ldots, x_{N,N}$ of the Hermitian matrices $X_N$ be
confined to a \emph{finite} interval $[a,b]$ and assume that the
empirical measure $N^{-1} \sum_j \delta(x - x_{j,N})$ converges
weakly to a measure with support in $[a,b]$. Under these conditions
the following is known.

Collins \cite{collins} differentiates the scaled logarithm of the
spherical integral $n$ times at zero to show that
\begin{equation}\label{eq:collins}
    \lim_{N\to\infty} N^{-1} \frac{d^n}{dk^n}\,\ln\,I_{X_N}(N k)
    \bigg\vert_{k=0} = (n-1)! \, c_n \;,
\end{equation}
i.e., he establishes convergence to the $n^\mathrm{th}$ free cumulant
(times a factorial). A stronger version of this result,
\begin{equation}\label{eq:g-m}
    \lim_{N\to\infty} \; N^{-1} \ln\, I_{X_N}(Nk) = \int_0^k
    R(k^\prime)\, dk^\prime = \sum_{n=1}^\infty \frac{c_n}{n}\, k^n\;,
\end{equation}
was proved by Guionnet and Maida \cite{GM} under the condition that
$k \in \mathbb{C}$ is small enough. (Notice that (\ref{eq:g-m})
implies (\ref{eq:collins}).) For $k$ real and large, however, the
authors of \cite{GM} obtain a \emph{different} behavior, separated
from the small-$k$ regime by a phase transition.

These results have a bearing on (\ref{eq:result-MZ}) for $p=1$, $q=0$
because the integral (\ref{eq:char-func}) can be done in two steps:
fixing some set of eigenvalues for $H$ we first do the integral over
$\mathrm{U}_N$ orbits -- that's precisely the spherical integral
(\ref{eq:HCIZ}) -- and afterwards we take the average over the
fluctuating eigenvalues. While it may seem puzzling at first sight
that the authors of \cite{GM} find a phase transition whereas we do
not, Section \ref{sect:match} explains that there is no contradiction
here. The assumption of eigenvalues strictly confined to an interval
$[a,b]$ would mean in our context that the confining potential $V(x)$
is infinitely high outside of $[a,b]$. In contradistinction, we
assume that $V$ is both uniformly convex and analytic. In the latter
setting the existence of a phase transition is ruled out on physical
and mathematical grounds. We in fact argue that the limit in
(\ref{eq:g-m}) is an entire function of $k \in \mathbb{C}\,$.

\subsubsection{Towards applications.}

By the general principles of invariant theory, the cha\-rac\-teristic
function $\Omega(K) = \Omega(g^{-1} K g)$ ($g \in \mathrm{GL}_N$)
lifts to a function $\widehat{\Omega}(Q)$ of the super\-matrix of
$\mathrm{GL}_N$-invariants
\begin{equation}
    Q = \left( \begin{array}{ll}
    \langle \widetilde{\varphi}_a \;, \varphi_{a^\prime} \rangle
    &\langle \widetilde{\varphi}_a \;, \psi_{b^\prime} \rangle \\
    \langle \widetilde{\psi}_b \;, \varphi_{a^\prime} \rangle
    &\langle \widetilde{\psi}_b \;, \psi_{b^\prime} \rangle
    \end{array} \right)_{ a,\,a^\prime=1,\ldots,p;\; b,\,b^\prime
    = 1, \ldots, q} \;.
\end{equation}
The angular brackets still mean contraction of the vector with the
co-vector. By the relation $\mathrm{Tr}\; K^n = \mathrm{STr}\, Q^n$
(where $\mathrm{STr}$ denotes the supertrace) it follows from
(\ref{eq:result-MZ}) that
\begin{equation}\label{eq:result-MZ2}
    \lim_{N \to \infty} N^{-1} \ln\, \widehat{\Omega}(N Q) =
    \sum_{n=1}^\infty \frac{c_n}{n}\; \mathrm{STr}\,Q^n\;.
\end{equation}
The lifted function $\widehat\Omega$ is input required by the
superbosonization method (Section \ref{sect:motiv}); in the past this
input was the link missing for applications. With
(\ref{eq:result-MZ2}) established, we now have at our disposal a
powerful new method for the treatment of random matrix problems. In
the present paper we illustrate the new method by demonstrating
universality for a random matrix model of stochastic scattering. In
particular, we point out that the pertinent large-$N$ saddle-point
equation for $Q$ is Voiculescu's equation $k^{-1} + R(k) = z$
generalized to the (super-)matrix case:
\begin{equation}\label{eq:speqn-V}
    Q^{-1} + R(Q) = z \;\mathrm{Id}_{p|q} \;.
\end{equation}

While the present paper addresses only the case of unitary symmetry,
our treatment is robust and readily extends to ensembles with
orthogonal or symplectic symmetry.

\subsection{Outline}

An outline of the contents of the paper is as follows. Section
\ref{sect:motiv} provides background and motivation by introducing
the characteristic function $\Omega(K)$ as a key object of the
supersymmetry method. For the special case of $K = \varphi \otimes
\widetilde{\varphi} = k \Pi$ with $k \in \mathbb{R}$ and a rank-one
projector $\Pi$, the large-$N$ asymptotics of $\Omega(NK)$ is
computed in Section \ref{sect:Dyson} using Dyson Coulomb gas methods.
Particular attention is paid to the fact that the asymptotics for
small and large $k$ match perfectly to give an answer which is smooth
as a function of $k$. The fermionic analog $K = \psi \otimes
\widetilde{\psi}$ with anti-commuting $\psi, \widetilde{\psi}$ is
treated by drawing on information from representation theory in
Section \ref{sect:FF}. By using standard perturbation theory in the
large-$N$ limit we then develop in Section \ref{sect:combinatorics} a
combinatorial description of the full superfunction $\Omega(NK)$. The
resulting formalism is applied to a model of stochastic scattering in
Section \ref{sect:scattering}. An outlook is given in Section
\ref{sect:last}.

\section{Review: supersymmetry method}\label{sect:motiv}

We begin the paper with a concise review of the supersymmetry method
and, in particular, of superbosonization. In this way, we shall
introduce and motivate the Fourier transform $\Omega(K) = \int
\mathrm{e}^{\mathrm{Tr} \, HK} d\mu_N(H)$ with $K$ given by
(\ref{eq:def-K}), which is a superfunction with symmetries and the
key object to be analyzed in the sequel.

\subsection{First steps}

Consider the Hermitian vector space $\mathbb{C}^N$ with its standard
Hermitian scalar product $h : \, \mathbb{C}^N \times \mathbb{C}^N \to
\mathbb{C}\,$, which determines a $\mathbb{C}$-antilinear bijection
\begin{equation}
    \dagger \, : \; \mathbb{C}^N \to (\mathbb{C}^N)^\ast \;,
    \quad v \mapsto v^\dagger := h(v , \cdot) \;,
\end{equation}
between $\mathbb{C}^N$ and its dual vector space $(\mathbb{C}^N
)^\ast$. Following standard physics conventions, we denote by the
same symbol (dagger) the operation $L \mapsto L^\dagger$ of taking
the Hermitian conjugate of a linear operator $L \in
\mathrm{End}(\mathbb{C}^N)$.

In the following the Hamiltonian $H$ will always be a random
Hermitian operator:
\begin{equation}
    \mathrm{End}(\mathbb{C}^N) \ni H = H^\dagger ,
\end{equation}
distributed according to some probability measure $\mu_N\,$. Our goal
is to study the spectral correlation functions which are defined as
averages with respect to $\mu_N\,$. For this purpose we consider the
characteristic polynomial $z \mapsto \mathrm{Det}(z-H)$ associated
with $H$. The supersymmetry method allows us to compute expectations
of products of ratios of such polynomials, and hence of products of
resolvent traces $\prod_a \mathrm{Tr}\,(z_a - H)^{-1}$, as follows.

Let us denote the canonical pairing $(\mathbb{C}^N)^\ast \otimes
\mathbb{C}^N \to \mathbb{C}$ (i.e., evaluation of a linear form
$\widetilde{\varphi} \in (\mathbb{C}^N)^\ast$ on a vector $\varphi
\in \mathbb{C}^N$) by
\begin{equation}
    \widetilde{\varphi} \otimes \varphi \mapsto
    \langle \widetilde{\varphi}, \varphi \rangle \;.
\end{equation}
With the resolvent operator $(z-H)^{-1}$ for $z \in \mathbb{C}
\setminus \mathbb{R}$ we associate a holomorphic function $\gamma\, :
(\mathbb{C}^N)^\ast \times \mathbb{C}^N \to\mathbb{C}$ by
\begin{equation}
    \gamma(\widetilde{\varphi},\varphi) = \mathrm{e}^{ - \langle
    \widetilde{\varphi} , \varphi \, z - H \varphi \rangle}\;.
\end{equation}
Now let $V_\mathbb{R}^z \subset (\mathbb{C}^N)^\ast \times
\mathbb{C}^N$ be the graph of the $\mathbb{R}$-linear mapping
\begin{equation}
    \mathbb{C}^N \to (\mathbb{C}^N)^\ast \;, \quad
    \varphi \mapsto - \mathrm{i} s \varphi^\dagger \;,
    \quad s := \mathrm{sign}(\mathfrak{Im}\, z) \in \{ \pm 1 \}\;.
\end{equation}
Thus $V_\mathbb{R}^z$ is the real vector space of all pairs
$(\widetilde{\varphi},\varphi) = (- \mathrm{i}s \varphi^\dagger,
\varphi)$ for $\varphi \in \mathbb{C}^N$. The Gaussian $\gamma$
decreases rapidly along $V_\mathbb{R}^z\,$. Indeed,
\begin{equation}
    \mathfrak{Re}\, \langle \widetilde{\varphi} , \varphi \,
    z - H \varphi \rangle \Big\vert_{V_\mathbb{R}^z} =
    \mathfrak{Re}\, h(\varphi,-\mathrm{i}s(z-H)\varphi)
    = |\mathfrak{Im}\, z| \; h(\varphi,\varphi) \geq 0 ,
\end{equation}
so we may integrate $\gamma$ along $V_\mathbb{R}^z\,$. By a standard
formula for Gaussian integrals we have
\begin{equation}\label{eq:det-0}
    \int_{V_\mathbb{R}^z} \mathrm{e}^{- \langle \widetilde{\varphi}
    , \varphi \, z - H \varphi \rangle} = \mathrm{Det}^{-1}(z-H) \;,
\end{equation}
where the integral is over $V_\mathbb{R}^z$ with ($\mathrm{i}
\mathbb{R} $-valued) Lebesgue measure normalized by the condition
$\int_{V_\mathbb{R}^z} \mathrm{e}^{ -\langle \widetilde{\varphi} ,
\varphi \rangle} = 1$. (This measure is not made explicit in our
notation.)

Expressing $\mathrm{Tr}\,(z-H)^{-1}$ as a logarithmic derivative,
\begin{equation}
    \mathrm{Tr}\,(z-H)^{-1} = \frac{d}{dz} \ln \mathrm{Det}(z-H)
    = \frac{\mathrm{Det}^\prime(z-H)}{\mathrm{Det}(z-H)} \;,
\end{equation}
we see that we need a Gaussian integration formula for
$\mathrm{Det}(z-H)$ (where $z \in \mathbb{C}$) in addition to that
for reciprocals $\mathrm{Det}^{ -1}(z-H)$. Such a formula can be had
by replacing commuting variables $\varphi$ by anticommuting variables
$\psi\,$; i.e., we view
\begin{equation}
    \langle \widetilde{\psi} , \psi\, z - H \psi \rangle
    \in \wedge^2( \mathbb{C}^N \oplus (\mathbb{C}^N)^\ast)
\end{equation}
as a quadratic element of the exterior algebra generated by the
direct sum $\mathbb{C}^N \oplus (\mathbb{C}^N)^\ast$. The precise
meaning is this. Let $\{ e_i \}$ be a basis of $\mathbb{C}^N$ and $\{
e^i \}$ be the dual basis of $(\mathbb{C}^N)^\ast$. Let $\ell : \,
\mathbb{C}^N \oplus (\mathbb{C}^N)^\ast \to \wedge(\mathbb{C}^N
\oplus (\mathbb{C}^N)^\ast)$ be the canonical embedding; or simply
put, view $\widetilde{\psi}_i \equiv \ell(e_i)$ and $\psi^i \equiv
\ell(e^i)$ as anticommuting variables or generators of the exterior
algebras $\wedge(\mathbb{C}^N)$ and $\wedge((\mathbb{C}^N)^\ast)$
respectively. Then we define $\langle \widetilde{\psi},\psi\,z - H
\psi\rangle$ to be the element of $\wedge^2( \mathbb{C}^N \oplus
(\mathbb{C}^N )^\ast)$ given by
\begin{equation}
    \langle \widetilde{\psi},\psi\,z - H \psi\rangle
    := \sum\nolimits_{i,j}
    \widetilde{\psi}_i\,\langle e^i ,(z-H)\, e_j \rangle \psi^j \;.
\end{equation}

By exponentiating this expression we get a Gaussian element in (the
even part of) the full exterior algebra:
\begin{equation}
    \mathrm{e}^{\langle \widetilde{\psi},\psi\,z - H \psi\rangle}
    \in \bigoplus\nolimits_{k=0}^N \wedge^{2k} ( \mathbb{C}^N
    \oplus (\mathbb{C}^N)^\ast) \;.
\end{equation}
The Berezin integral $f \mapsto \int f$ for $f \in \wedge(
\mathbb{C}^N \oplus (\mathbb{C}^N)^\ast)$ is, by definition,
projection on the one-dimensional subspace $\wedge^{2N}( \mathbb{C}^N
\oplus (\mathbb{C}^N)^\ast)$ of top degree. In the case of our
Gaussian integrand the result of this projection is known to be
proportional to the determinant of the operator $z-H$. We normalize
the Berezin integral in such a way that the constant of
proportionality is unity:
\begin{equation}\label{eq:det-1}
    \int \mathrm{e}^{\langle \widetilde{\psi}, \psi
    \, z - H \psi \rangle} = \mathrm{Det}(z-H) \;.
\end{equation}
To summarize the above, we have two Gaussian integration formulas:
Eq.\ (\ref{eq:det-1}) for the secular determinant $\mathrm{Det} (z -
H)$, and Eq.\ (\ref{eq:det-0}) for its reciprocal.

By multiplying these formulas, averaging the result with the given
probability density $d\mu_N\,$, and interchanging the order of
integrations, we obtain
\begin{equation}\label{eq:ratio}
    \int \frac{\mathrm{Det}(w_1 - H)}{\mathrm{Det}(w_0 - H)}\,
    d\mu_N(H) = \int_{V_\mathbb{R}^{w_0}} \Omega(\varphi
    \otimes \widetilde{\varphi} + \psi \otimes \widetilde{\psi})\,
    \mathrm{e}^{- w_0 \langle\widetilde{\varphi} , \varphi
    \rangle + w_1 \langle \widetilde{\psi}, \psi \rangle}
\end{equation}
where $w_1 \in \mathbb{C}\,$, $w_0 \in \mathbb{C} \setminus
\mathbb{R}\,$, and $\Omega$ is the characteristic function
\begin{equation}
    \Omega(K) = \int \mathrm{e}^{\mathrm{Tr}\, HK} d\mu_N(H)\;.
\end{equation}
Note that the formula (\ref{eq:ratio}) requires the evaluation of
$\Omega$ for
\begin{equation}
    K = \varphi \otimes \widetilde{\varphi} + \psi \otimes
    \widetilde{\psi} = \sum\nolimits_{i,j=1}^N \big(
    \varphi^i \widetilde{\varphi}_j + \psi^i \widetilde{\psi}_j
    \big) e_i \, \langle e^j , \cdot \rangle \;,
\end{equation}
where $\varphi^i : \, \mathbb{C}^N \to \mathbb{C}$ and
$\widetilde{\varphi}_j : \, (\mathbb{C}^N)^\ast \to \mathbb{C}$ are
the linear coordinates associated with the bases $\{ e_i \}$ of
$\mathbb{C}^N$ and $\{ e^j \}$ of $(\mathbb{C}^N)^\ast$. Thus the
operator $K$ is an endomorphism of $\mathbb{C}^N$ with coefficients
in the tensor product
\begin{equation}
    \mathrm{S}( \mathbb{C}^N \oplus (\mathbb{C}^N)^\ast)
    \otimes \wedge( \mathbb{C}^N \oplus (\mathbb{C}^N)^\ast)
\end{equation}
of the symmetric and exterior algebras of $\mathbb{C}^N \oplus
(\mathbb{C}^N)^\ast$. It will be important that the numerical part of
$K = \varphi \otimes \widetilde{\varphi} + \psi \otimes \widetilde
{\psi}$ has finite rank.

The relation (\ref{eq:ratio}) transfers the integral over $N\times N$
random matrices $H$ to an integral over the variables $\varphi, \psi$
constituting the bilinear $K$. This transfer will be a step forward
if we can calculate the function $\Omega(K)$ or, at least, gather
enough information about it. For the case of a Gaussian probability
measure $\mu_N\,$, the Fourier-Laplace transform $\Omega(K)$ is also
Gaussian. The supersymmetry formalism then takes its course and
delivers results quickly. However, using the traditional version of
the supersymmetry method one did not know how to proceed in the
general case of non-Gaussian $\mu_N\,$.

\subsection{Superbosonization}

One way to proceed, as we shall now review, is to make a symmetry
assumption about $\Omega(K)$. Let $V := (\mathbb{C}^N)^\ast \oplus
\mathbb{C}^N$. Then $\mathrm{e}^{\mathrm{Tr}\, HK}$ for $K = \varphi
\otimes \widetilde{\varphi} + \psi \otimes \widetilde {\psi}$ is a
superfunction $f : \, V \to \wedge(V^\ast)$, and by making the
identifications $\psi^i \equiv d\varphi^i$ and $\widetilde{\psi}_i
\equiv d\widetilde{\varphi}_i$ we may regard $f = \mathrm{e}^{
\mathrm{Tr}\, HK}$ as a holomorphic differential form on the complex
vector space $V$. Let now $\mu_N$ be invariant under conjugation $H
\mapsto g H g^{-1}$ by the elements $g$ of some compact group $G$.
Then by integrating $f$ against $d\mu_N$ we obtain a differential
form $\Omega:\, V \to \wedge(V^\ast)$ which is $G$-equivariant, i.e.,
\begin{equation}
    \Omega(v) = g \Omega(g v) \;,
\end{equation}
where $(g,v) \mapsto g v$ and $(g,\Omega) \mapsto g \Omega$ are the
natural $G$-actions on $V$ resp. $\wedge(V^\ast)$. Later we will
write this equivariance property more intuitively as $\Omega(K) =
\Omega(g K g^{-1})$.

In such a setting, the superbosonization method offers a reduction
step which is available \cite{LSZ-D2} for the classical Lie groups $G
= \mathrm{U}_N\,$, $\mathrm{O}_N\,$, and $\mathrm{USp}_N\,$. For each
of these groups, the algebra of $G$-equivariant differential forms on
$V$ is generated by (the dual of) the $\mathbb{Z}_2 $-graded vector
space $W = W_0 \oplus W_1$ of quadratic $G$-invariants
\cite{howeRCIT}. After lifting the differential form $\Omega : \, V
\to \wedge(V^\ast)$ to a superfunction $\widehat\Omega : \, W_0 \to
\wedge(W_1^\ast)$, the step of `superbosonization' transfers the
integral on the r.h.s.\ of (\ref{eq:ratio}) to an integral of the
lifted superfunction over a (low-dimensional) Riemannian symmetric
superspace.

We now reproduce from \cite{LSZ-D2} the details for the case of $G =
\mathrm{U}_N\,$, with a few notational adjustments to fit the present
situation. By immediate generalization of (\ref{eq:ratio}) we have
\begin{eqnarray}
    \int \frac{\prod_{b=1}^q \mathrm{Det}(w_{1,\,b}-H)}{\prod_{a=1}^p
    \mathrm{Det}(w_{0,\,a}-H)}\; d\mu_N(H) = \cr \int \Omega \Big(
    \sum\nolimits_a \varphi_a \langle \widetilde{\varphi}_a \,,
    \cdot\rangle + \sum\nolimits_b \psi_b \langle\widetilde{\psi}_b
    \, ,\cdot \rangle \Big) \, \mathrm{e}^{- \sum w_{0,\,a}\, \langle
    \widetilde{\varphi}_a , \varphi_a \rangle + \sum {w_{1,\,b}}\,
    \langle \widetilde{\psi}_b , \psi_b \rangle} ,
\end{eqnarray}
where the $\varphi$-integral is over the real subspace
\begin{equation}
    \widetilde{\varphi}_a = - \mathrm{i} s_a \varphi_a^\dagger
    \;,\quad s_a = \mathrm{sign}\, (\mathfrak{Im}\, w_{0,\,a})\;,
    \quad a = 1, \ldots, p \;.
\end{equation}
While intending to specialize to the case of $p = q$ later, we here
describe the general case $p \not= q$ for a clear exposition of the
formalism.

To simplify the notation, it is convenient to regard the vectors
$\varphi_1 , \ldots, \varphi_p$ as the components of a linear mapping
$\varphi$ from $\mathbb{C}^p$ to $\mathbb{C}^N:$
\begin{equation}
    \varphi := (\varphi_1 , \ldots, \varphi_p) \in
    \mathrm{Hom}(\mathbb{C}^p , \mathbb{C}^N) \;.
\end{equation}
Similarly, we view $\widetilde{\varphi} := (\widetilde{\varphi}_1 ,
\ldots, \widetilde{\varphi}_p)$ as a linear mapping
\begin{equation}
    \widetilde{\varphi} \in \mathrm{Hom}(\mathbb{C}^N ,\mathbb{C}^p)\;.
\end{equation}
Using the same conventions on the anti-commuting side, we write our
integral as
\begin{eqnarray}
    \int \frac{\prod_{b=1}^q \mathrm{Det}(w_{1,\,b}-H)} {\prod_{a=1}^p
    \mathrm{Det}(w_{0,\,a}-H)}\; d\mu_N(H)\cr = \int \Omega\big( \varphi
    \widetilde{\varphi} + \psi \widetilde{\psi} \big)\,\mathrm{e}^{-
    \mathrm{Tr}\, (\varphi\, w_0 \widetilde{\varphi} + \psi w_1
    \widetilde{\psi})} \;. \label{eq:int-formula}
\end{eqnarray}
Here $\mathrm{Tr} = \mathrm{Tr}_{\mathbb{C}^N}\,$, and $w_0 = \mathrm
{diag}(w_{0,1}, \ldots, w_{0,p})$, $w_1 = \mathrm{diag}( w_{1,1},
\ldots, w_{1,q})$ are diagonal operators. The integral is over
\begin{equation}
    \widetilde{\varphi} = - \mathrm{i} s \varphi^\dagger \;,
    \quad s = \mathrm{diag}(s_1, \ldots, s_p) \;.
\end{equation}

It is evident that the integrand on the r.h.s.\ of
(\ref{eq:int-formula}) has the invariance property
\begin{equation}
    f(\varphi,\widetilde{\varphi},\psi,\widetilde{\psi})
    = f(g \varphi,\widetilde{\varphi} g^{-1}, g \psi,
    \widetilde{\psi} g^{-1})
\end{equation}
for $g \in \mathrm{U}_N$ and hence, by holomorphic continuation, for
$g \in \mathrm{GL}_N\,$. This implies that there exists \cite{LSZ-D2}
a (lifted) function $\widehat{f}(Q)$ of a supermatrix $Q = \left(
\begin{array}{ll} x &\sigma\\ \tau &y \end{array} \right)$ such that
\begin{equation}
    \widehat{f} \left( \begin{array}{ll} \widetilde{\varphi} \varphi
    &\widetilde{\varphi} \psi \\ \widetilde{\psi} \varphi
    &\widetilde{\psi} \psi \end{array}\right) =
    f(\varphi,\widetilde{\varphi},\psi,\widetilde{\psi}) \;.
\end{equation}
Precisely speaking, $\widehat{f} : \, W_0 \to \wedge(W_1^\ast)$ is a
function on the $\mathbb{Z}_2$-even vector space $W_0 = \mathrm{End}
(\mathbb{C}^p)\oplus \mathrm{End}(\mathbb{C}^q)$ (the diagonal blocks
of $Q$) with values in the exterior algebra of the dual of the
$\mathbb{Z}_2$-odd vector space $W_1 = \mathrm{Hom} (\mathbb{C}^p ,
\mathbb{C}^q) \oplus \mathrm{Hom}(\mathbb{C}^q , \mathbb{C}^p)$ (the
off-diagonal blocks of $Q$). The lift $\widehat{f}(Q)$ turns into the
given function $f(\varphi, \widetilde{\varphi}, \psi, \widetilde
{\psi})$ upon substituting for $Q$ the quadratic
$\mathrm{GL}_N$-invariants,
\begin{equation}
    Q = \left( \begin{array}{ll} x &\sigma\\ \tau &y \end{array}
    \right) \to \left(\begin{array}{ll} \widetilde{\varphi} \varphi
    &\widetilde{\varphi} \psi\\ \widetilde{\psi}\varphi
    &\widetilde{\psi}\psi\end{array} \right) \;.
\end{equation}

The ensuing step of superbosonization exploits the $\mathrm{GL}_N
$-symmetry of the integrand to implement a reduction: it transfers
the integral over the variables $\varphi, \psi$ to an integral over
the supermatrices $Q:$
\begin{equation}\label{eq:supbos}
    \int f(\varphi,\widetilde{\varphi},\psi,\widetilde{\psi}) =
    c_{p,\,q} \int DQ\; \mathrm{SDet}^N(Q) \, \widehat{f}(Q) \;.
\end{equation}
The integral on the right-hand side is over
\begin{equation}
    \mathrm{H}_p^s \times \mathrm{U}_q \subset \mathrm{End}(\mathbb{C}^p)
    \times \mathrm{End}(\mathbb{C}^q) \;,
\end{equation}
where $\mathrm{U}_q \equiv \mathrm{U}(\mathbb{C}^q)$ is the unitary
group of $\mathbb{C}^q$ and
\begin{equation}\label{eq:Hps}
    \mathrm{H}_p^s := \{ -\mathrm{i}s M \mid M = M^\dagger > 0\}
\end{equation}
is a space isomorphic to the positive Hermitian $p \times p$ matrices
(replacing the quadratic $\mathrm{U}_N$-invariant $\varphi^\dagger
\varphi$). With a natural choice \cite{LSZ-D2} of normalization for
the Berezin integration form $DQ$ the normalization factor is $c_{p,
\,q} = \mathrm{vol}(\mathrm{U}_n) / \mathrm{vol}(\mathrm{U}_{n-p+
q})$. In the important special case of $p = q$ the Berezin
integration form $DQ$ is simply the product of differentials times
the product of derivatives:
\begin{equation}\label{eq:DQ}
    DQ \propto \prod_{a,\,a^\prime} dx_{a a^\prime}\;
    \prod_{b,\,b^\prime} dy_{b b^\prime} \; \prod_{a,\,b}
    \frac{\partial^2}{\partial\sigma_{ab} \partial\tau_{ba}} \;.
\end{equation}
Finally, it should be stressed that the formula (\ref{eq:supbos}) is
valid if and only if $N \geq p$.

Application of the superbosonization formula (\ref{eq:supbos}) to
Eq.\ (\ref{eq:int-formula}) yields the identity
\begin{eqnarray}
    \int \frac{\prod_{b=1}^q \mathrm{Det}(w_{1,\,b} - H)}
    {\prod_{a=1}^p \mathrm{Det}(w_{0,\,a} - H)} \; d\mu_N(H)\cr
    = c_{p,\,q} \int DQ\; \mathrm{SDet}^N(Q) \, \widehat{\Omega}(Q)\,
    \mathrm{e}^{-\mathrm{STr}\; w Q} \;, \label{eq:supbos2}
\end{eqnarray}
where $\mathrm{STr}\, w Q \equiv \mathrm{Tr}_{\mathbb{C}^p} (w_0 Q) -
\mathrm{Tr}_{\mathbb{C}^q} (w_1 Q)$. The function
$\widehat{\Omega}(Q)$ is a lift of the characteristic function
$\Omega(K)$ for $K = \varphi \widetilde{\varphi} + \psi
\widetilde{\psi}$.

In view of the result (\ref{eq:supbos2}) our short-term goals should
now be well motivated: the key object to understand is the lifted
characteristic (super-)function $\widehat{\Omega}(Q)$. If we can
control $\widehat\Omega\,$, results for the level correlation
functions will follow from a large-$N$ asymptotic saddle analysis of
the $Q$-integral.

\section{Coulomb gas argument}
\label{sect:Dyson}

Based on what is called the Dyson Coulomb gas, we are now going to
study $\Omega(K)$ for the rank-one case $K = k\Pi$ with $k \in
\mathbb{R}$ and $\Pi$ the projector on a one-dimensional subspace of
$\mathbb{C}^N$. A related situation has been investigated in the work
of P.\ Zinn-Justin \cite{pzj-CMP,pzj-PRE}, Collins \cite{collins},
and Guionnet \& Maida \cite{GM}; we will comment on the literature as
we go along. We begin by reviewing some basic material.

\subsection{Voiculescu $R$-transform}

As before, let $\mu_N$ be a probability measure (with or without
invariance properties) for Hermitian $N \times N$ matrices $H$. Let
\begin{equation}\label{eq:moments}
    m_{n,\,N} := N^{-1} \int \mathrm{Tr}\,(H^n)\, d\mu_N(H)
\end{equation}
denote the $n^\mathrm{th}$ moment of $\mu_N\,$. We shall assume that
$m_{n,\,N}$ has a limit,
\begin{equation}
    m_n := \lim_{N\to\infty} m_{n,\,N} \;,
\end{equation}
and consider a generating function $z \mapsto g(z)$ for these moments
at $N = \infty:$
\begin{equation}\label{eq:g(z)-series}
    g(z) := \sum_{n=0}^\infty m_n \, z^{-n-1} \;.
\end{equation}
This series (when it converges), or rather its analytic continuation
\begin{equation}
    g(z) = \lim_{N \to \infty} N^{-1}
    \int \mathrm{Tr}\,(z-H)^{-1} d\mu_N(H) \qquad
    (z \in \mathbb{C} \setminus \mathbb{R})\;,
\end{equation}
is the Cauchy transform
\begin{equation}\label{eq:Cauchy}
    g(z) = \int_\mathbb{R} \frac{d\nu(x)}{z-x} \;, \qquad
    d\nu(x) = \pi^{-1} \lim_{\epsilon \to 0+} \mathfrak{Im}\,
    g(x-\mathrm{i}\epsilon)\, dx \;,
\end{equation}
of the large-$N$ limit of the so-called density of states $\nu$. Let
us now assume that $\nu$ has compact support. Then there exists a
number $r > 0$ such that for $z \in \mathbb{C}$ with $|z| > r$ the
power series (\ref{eq:g(z)-series}) converges, the derivative
$g^\prime(z)$ does not vanish, and hence by the implicit function
theorem the function $z \mapsto g(z)$ has a local inverse. Because
the expansion of $g(z)$ around $z = \infty$ begins as $k \equiv g(z)
= z^{-1} + \ldots\,$, the power series for the inverse function will
likewise begin as $z = k^{-1}$ plus corrections. These considerations
lead to Voiculescu's definition of the $R$-transform \cite{voicul1}:
\begin{equation}\label{eq:Voiculescu}
    g(z) = k \Longleftrightarrow k^{-1} + R(k) = z \;, \qquad
    R(k) = \sum\nolimits_{n=1}^\infty c_n\, k^{n-1} \;.
\end{equation}
Thus the $R$-transform $k \mapsto R(k)$ is the function inverse to $z
\mapsto g(z)$ with the pole $k^{-1}$ subtracted. Under the assumption
of compact support for $\nu$ the power series for $R(k)$ converges
for sufficiently small $k\,$. The coefficients $c_n$ are called
\emph{free cumulants}.

Freeness of random variables is defined in an algebraic way
\cite{VDN} which will not be reviewed here. Suffice it to say that
two random matrices $A$, $B$ are free (in the limit $N = \infty$) if
the probability law of $A + B$ remains unchanged under conjugation $B
\mapsto U^\dagger B\, U$ by any $U \in \mathrm{U}_\infty\,$.
Voiculescu proved \cite{voicul1} that the $R$-transform is linear for
free convolution, i.e., if $A$, $B$ are free [with $R$-transforms
$R_A(k)$ resp.\ $R_B(k)$], then $R_{A+B}(k) = R_A(k) + R_B(k)$. This
linearity property parallels the fact that the cumulants of a sum of
commutative random variables equal the sum of the cumulants, and it
leads to the expectation that the $R$-transform is closely related to
the logarithm of the Fourier transform of $\mu_N \,$, $N \to \infty$.

Let us mention in passing that, since $g(z)$ is also known as the
Green's function, physicists by Zee's fancy \cite{zee} sometimes call
$b(k) := k^{-1} + R(k)$ the Blue's function.

\subsubsection{Examples.}

The Gaussian measure $\mu_N$ with density $d\mu_N(H) \propto
\mathrm{e}^{-\frac{N}{2} \mathrm{Tr}\,H^2} dH$ is called the Gaussian
Unitary Ensemble (GUE) with $c_2 = 1$. For this measure one has $R(k)
= k$ and solving the equation $z = k^{-1} + R(k) = k^{-1} + k$ for $k
= g(z)$ one finds
\begin{equation}
    g(z) = {\textstyle{\frac{1}{2}}} \big(z \pm \sqrt{z^2-4}\,\big) ,
\end{equation}
which gives Wigner's semicircle law $d\nu(x) = (2\pi)^{-1} \sqrt{4 -
x^2}\; dx\,$.

Another example (taken from recent work \cite{LSZ-C3} by Lueck,
Sommers and one of the authors, on the energy correlations of a
random matrix model for disordered bosons) is this. Consider
\begin{equation}
    R(k) = \frac{k}{1-k^2} = \sum_{n=1}^\infty k^{2n-1} \;.
\end{equation}
Thus all odd free cumulants vanish and the even free cumulants are
all equal to unity. Solving Voiculescu's equation $k^{-1} + R(k) = z$
for $k = g(z)$ one obtains
\begin{equation}
    g(z) = \mathrm{i} \left( \sqrt{\frac{1}{27} - \frac{1}{4z^2}} -
    \frac{\mathrm{i}}{2z} \right)^{1/3}
    - \mathrm{i} \left( \sqrt{\frac{1}{27} - \frac{1}{4z^2}} +
    \frac{\mathrm{i}}{2z} \right)^{1/3} \;.
\end{equation}
The density of states of this example has compact support but is
unbounded due to an inverse cube root singularity $|x|^{-1/3}$ at $x
= 0\,$.

\subsection{Eigenvalue reduction of $\Omega$}

We now specialize to the case of a $\mathrm{U}_N$-invariant
probability measure for Hermitian operators $H$ with density $d\mu_N
(H) = \mathrm{e}^{-N\, \mathrm{Tr}\, V(H)} dH$. The goal here is to
relate the $R$-transform $R(k)$ to the large-$N$ limit of the
characteristic function (\ref{eq:char-func}) for $K = k \Pi$ with
$\Pi$ a rank-one projector. Our approach will be similar to that of
P.\ Zinn-Justin \cite{pzj-PRE} based on the
Harish-Chandra-Itzykson-Zuber integral.

We start by diagonalizing the Hamiltonian $H$ by a unitary
transformation:
\begin{equation}
    H = g^{-1} X g\;,\quad X = \mathrm{diag}
    (x_1,\ldots,x_N)\;, \quad g \in \mathrm{U}_N\;.
\end{equation}
Recalling that the Jacobian $J(X)$ associated with this
transformation is the square of the Vandermonde determinant:
\begin{equation}
    J(X) = \prod_{i < j} (x_i - x_j)^2 \;,
\end{equation}
we cast the expression for the characteristic function in the form
\begin{equation}\label{eq:mrz-x0}
    \Omega(K) = C_N \int_{\mathbb{R}^N} \left(\int_{\mathrm{U}_N}
    \mathrm{e}^{\mathrm{Tr}\,(X g K g^{-1})} dg \right)
    \mathrm{e}^{-N\,\mathrm{Tr}\; V(X)} J(X)\, d^N x \;,
\end{equation}
where $dg$ is a Haar measure for $\mathrm{U}_N\,$. The normalization
constant $C_N$ is determined by the condition $\Omega(0) = 1$.

Now let $K \equiv k \Pi$ where $k \in \mathbb{R}$ and $\Pi$ is the
orthogonal projector on some (fixed) complex line in $\mathbb{C}^N$.
We are then faced with the inner integral
\begin{equation}
    \int_{\mathrm{U}_N} \mathrm{e}^{k\,
    \mathrm{Tr}\,(X g \Pi g^{-1})} dg \;.
\end{equation}
The integrand depends on $g \in \mathrm{U}_N$ only through the
projector $\Pi$ conjugated by $g$, and the set of all these
projectors $g \Pi g^{-1}$ is in bijection with the projective space
$\mathbb{C}P^{N -1} \simeq \mathrm{U}_N / (\mathrm{U}_1 \times
\mathrm{U}_{N-1})$ of complex lines in $\mathbb{C}^N$. By
parametrizing $g \Pi g^{-1}$ in the eigenbasis of $H$ as $(g \Pi
g^{-1})_{ij} = u_i\, \bar{u}_j$ with $\sum_{j=1}^N |u_j|^2 = 1$, we
reduce our integral to
\begin{equation}\label{eq:mrz-x1a}
    \int_{\mathrm{U}_N} \mathrm{e}^{k\, \mathrm{Tr}\,
    (X g \Pi g^{-1})} dg  = \int_{\mathbb{C}P^{N-1}}
    \mathrm{e}^{k \sum_{j=1}^N x_j\, |u_j|^2} du \;,
\end{equation}
where $du$ is a $\mathrm{U}_N$-invariant measure for
$\mathbb{C}P^{N-1}$.

Now $\mathbb{C}P^{N-1}$ is a K\"ahler manifold with $\mathrm{U}_N
$-invariant Riemannian geometry, and the function
\begin{equation}
    \mu \, : \; \mathbb{C}P^{N-1} \to \mathrm{Lie}\, \mathrm{U}_N \;,
    \quad g \cdot (\mathrm{U}_1 \times \mathrm{U}_{N-1}) \mapsto
    \mathrm{i} g \Pi g^{-1},
\end{equation}
is a momentum mapping \cite{BGV}. We observe that the expression $k\,
\mathrm{Tr} \,(X g \Pi g^{-1})$ in the exponent of our integrand is
obtained by contracting $\mu$ with the Lie algebra element
$-\mathrm{i}k X \in \mathrm{Lie}\, \mathrm{U}_N\,$. It follows that
the integral is governed by the Duistermaat-Heckman localization
principle \cite{BGV}. In other words, the integral can be computed
exactly by performing the stationary-phase approximation (including
the Gaussian fluctuations) for each of its critical points and
summing the contributions.

There are $N$ critical points; these are the points where $g \Pi
g^{-1}$ is diagonal (with one diagonal element equal to unity and all
others equal to zero). By computing the contribution from each point
and taking the sum, we get
\begin{equation}\label{eq:DH}
    \int_{\mathbb{C}P^{N-1}} \mathrm{e}^{k \sum_{j=1}^N x_j\,|u_j|^2}
    du= C_{1,\,N}\,k^{-(N-1)}\sum_{i=1}^N \frac{\mathrm{e}^{k x_i}}
    {\prod_{j (\not= i)} (x_i - x_j)}\;,
\end{equation}
with some $N$-dependent constant $C_{1,\,N}\,$.

\subsection{Coulomb gas}\label{sect:Coulomb}

Based on the exact expressions (\ref{eq:mrz-x0})--(\ref{eq:DH}), our
goal is to compute the large-$N$ asymptotics of $N^{-1} \ln
\Omega(Nk\Pi)$. We begin by recalling \cite{deift} that, governed by
\begin{equation}
    \prod_{i<j} (x_i - x_j)^2 \prod_l \mathrm{e}^{-N\,V(x_l)}
    dx_l \;,
\end{equation}
the eigenvalues $x_1, \ldots, x_N$ distribute for $N \to \infty$
according to the equilibrium measure, $\nu$, which is determined by
minimizing Dyson's Coulomb gas energy functional:
\begin{equation}
    N^2 \int V(x)\,d\nu(x) - N^2 \int\!\!\!\int \ln|x-y|\,d\nu(x)\,
    d\nu(y)\;,
\end{equation}
the energy of a gas or fluid of charged particles subject to a
confining potential $N\,V$ and mutual repulsion by (the
two-dimensional form of) Coulomb's law. The Euler-Lagrange equation
for the Coulomb gas energy functional reads
\begin{equation}\label{eq:Euler}
    V(x) - 2 \int\ln |x-y| \; d\nu(y) + \ell = 0
    \qquad (x \in \mathrm{supp}\; \nu) \;,
\end{equation}
where $\ell$ is a Lagrange multiplier for the normalization
constraint $\int d\nu(x) = 1$. By differentiating once with respect
to $x$ one obtains
\begin{equation}\label{eq:V-from-g}
    V^\prime(x) = 2\, \mathrm{P.V.} \int \frac{d\nu(y)}
    {x-y} \qquad (x \in \mathrm{supp}\; \nu) \;.
\end{equation}
Physically speaking, this condition means that the total force
vanishes in the state of equilibrium inside the fluid.

The task of determining the measure $\nu$ from Eq.\ (\ref{eq:Euler})
can be formulated and solved as a Riemann-Hilbert problem
\cite{deift}. It is known that the solution is unique and corresponds
to a minimum of the energy (hence a maximum of the integrand).

From now on we shall simplify our work by taking the confining
potential $V$ of the probability measure $\mu_N$ to be \emph{convex}.
This assumption ensures that the large-$N$ density of states is
supported on a single interval: $\mathrm{supp}\; \nu = [a,b]$.

Next recall the definition (\ref{eq:Cauchy}) of the Cauchy transform
$g(z)$. Denoting by $g_\pm$ the two limits
\begin{equation}
    g_\pm(x) = \lim_{\varepsilon\to 0+} g(x\pm\mathrm{i}\varepsilon)\;,
\end{equation}
which $g(z)$ takes on approaching $x \in [a,b]$ from the upper or
lower half of the complex plane, one rewrites the relation
(\ref{eq:V-from-g}) as
\begin{equation}
    V^\prime(x) = g_+(x) + g_-(x) \;.
\end{equation}

As it stands, this equation holds only for $x \in [a,b] \subset
\mathbb{R}\,$. However, from general theory \cite{DKMVZ} one knows
that $g_{\pm} (x)$ are the two branches of a double-valued
complex-analytic function $z \mapsto (g(z),h(z))$ evaluated at $z =
x\,$. Thus by the principle of analytic continuation we have, for all
$z \in \mathbb{C} \setminus \{a , b\}$,
\begin{equation}\label{eq:ana-cont}
    V^\prime(z) = g(z) + h(z) \;.
\end{equation}
Note that $g(a) = h(a) = g_+(a) = g_-(a)$ and $g(b) = h(b) = g_+(b) =
g_-(b)$.

We will also make use of the integrated form of Eq.\
(\ref{eq:ana-cont}). For $z \in \mathbb{C} \setminus (-\infty,b]$ let
\begin{equation}
    G(z) := \int_a^b \ln (z - x) \, d\nu(x) \;,
\end{equation}
where the principal branch of the logarithm is assumed, giving $G(x)
\in \mathbb{R}$ for $x \in (b,\infty)$. Then let $H : \, \mathbb{C}
\setminus (-\infty,b] \to \mathbb{C}$ be defined by the equation
\begin{equation}\label{eq:mrz-i1}
    V(z) = G(z) + H(z) - \ell \;.
\end{equation}
We observe that $G^\prime(z) = g(z)$ and hence $H^\prime(z) = h(z)$.
Moreover, by combining (\ref{eq:mrz-i1}) with the Euler-Lagrange
equation (\ref{eq:Euler}) we have
\begin{equation}\label{eq:G=H}
    H(b) = V(b) - G(b) + \ell = V(b) - \int_a^b \ln |b - x| \;
    d\nu(x) + \ell = G(b) \;.
\end{equation}
By the same reasoning, $H(a) = G(a)$.

\subsection{Asymptotics for $k$ small}\label{sect:small-k}

In this subsection, we take the absolute value of the real number $k$
to be small. In order to compute the large-$N$ asymptotics of $\ln
\Omega(N k\Pi)$ for this situation, we modify the exact integral
representation (\ref{eq:mrz-x0})--(\ref{eq:DH}) by expressing the
right-hand side of (\ref{eq:DH}) as a complex contour integral:
\begin{equation}\label{eq:CIR}
    \sum_{i=1}^N \frac{\mathrm{e}^{k x_i}} {\prod_{j (\not= i)}(x_i
    - x_j)} = \frac{1}{2\pi\,\mathrm{i}} \oint\limits_{\mathcal{C}_x}
    \frac{\mathrm{e}^{kz}\, dz}{\prod_{j=1}^N (z-x_j)} \;,
\end{equation}
where the contour $\mathcal{C}_x$ loops around the set of points
$x_1, \ldots, x_N\,$. We thus obtain
\begin{eqnarray}
    \Omega(Nk\Pi) = C_{2,\,N}\, k^{-N+1} \times \cr
    \times \int_{\mathbb{R}^N} \left(\oint_{\mathcal{C}_x}
    \frac{\mathrm{e}^{Nkz}\, dz}{\prod_{l=1}^N (z-x_l)} \right)
    \prod_{i<j} (x_i-x_j)^2 \prod_l \mathrm{e}^{-N\,V(x_l)}dx_l\;.
\end{eqnarray}

Because the Coulomb gas energy is of order $N^2$, which is large
compared to the perturbation due to the contour integral and here in
particular the `external electric field' term $N k z\,$, we expect
the density of the fluid of charges $x_1, \ldots, x_N$ to remain the
same in the limit $N \to \infty$. (This will turn out to be fully
correct as long as $k$ does not exceed a critical value.) Thus the
charges $x_1, \ldots, x_N$ are still expected to distribute (for $N
\to \infty$) according to the equilibrium measure $\nu$ with support
$[a,b]$.

Taking this fact for granted, we fix a contour $\mathcal{C}$
encircling $\mathrm{supp}\, \nu = [a,b]$ and interchange the integral
over $\{ x_1, \ldots, x_N\}$ with the contour integral. Then, by
taking the logarithm and passing to the large-$N$ limit we arrive at
\begin{eqnarray}
    \omega(k) := \lim_{N\to\infty} N^{-1} \ln \Omega(Nk\Pi) \cr
    = \gamma + \lim_{N\to\infty} N^{-1} \ln \oint_{\mathcal{C}}
    \mathrm{e}^{Nkz - N \int_a^b \ln\,(kz-kx)\, d\nu(x)} dz\;,
    \label{eq:int-small-k}
\end{eqnarray}
with some number $\gamma$ which remains unknown for the moment, as we
did not keep track of the overall normalization constant. Note that
$\omega(0) = 0$ from $\Omega(0) = 1$.

The integral (\ref{eq:int-small-k}) for $N \to \infty$ is computable
by saddle analysis and the method of steepest descent. We first look
for the critical points of the integrand. The condition for $z \equiv
z_0$ to be an extremum is
\begin{equation}
    k = \int_a^b \frac{d\nu(x)}{z_0 - x} = g(z_0) \;.
\end{equation}
When does this equation have a solution for $z_0 \in \mathbb{R}$? To
find the answer, notice that the function $g(x)$ for $x \in
\mathbb{R} \setminus [a,b]$ is monotonically decreasing:
\begin{equation}
    g^\prime(x) = - \int_a^b \frac{d\nu(y)}{(x-y)^2} < 0
    \qquad (x \notin [a,b]) \;.
\end{equation}
We therefore have the inequality $g(b) > g(\infty) = 0 > g(a)$ and
the function
\begin{equation}
    g \, : \; \mathbb{R} \setminus (a,b) \to [g(a),g(b)]
\end{equation}
is a bijection. Thus for any $k \in [g(a),g(b)]$ there exists a
unique solution
\begin{equation}
    z_0 = g^{-1}(k)
\end{equation}
of the equation $k = g(z_0)$. Note that $z_0 \in \mathbb{R} \setminus
(a,b)$.

In the following let $k$ be fixed, with $g(a) < k < g(b)$. To
evaluate the integral (\ref{eq:int-small-k}) by steepest descent, we
deform the contour $\mathcal{C}$ for $z$ into the axis $g^{-1}(k) +
\mathrm{i}\mathbb{R}$ parallel to the imaginary $z$-axis. Because
$g^\prime(x) < 0$ for $x \in \mathbb{R} \setminus [a,b]$, the saddle
at $z_0 = g^{-1}(k)$ is a local minimum of the integrand evaluated
along the real axis, but is a local maximum of the integrand on the
axis $g^{-1} (k)+ \mathrm{i}\mathbb{R}\,$. Thus the path of steepest
descent leads across the saddle $z_0 = g^{-1}(k)$ in the direction of
$\mathrm{i} \mathbb{R}\,$. Steepest-descent evaluation of the
integral then yields
\begin{equation}\label{eq:small-k}
    \omega(k) = - 1 + k\, g^{-1}(k) - \int_a^b \ln
    \big(k\,g^{-1}(k)- k\,x \big) \, d\nu(x) \;.
\end{equation}
Notice that since $k\, g^{-1}(k) \to 1$ as $k \to 0\,$, this
satisfies the required normalization condition $\omega(0) = 0$ by
insertion of the additive constant $\gamma = - 1$. For later use, we
write our result in the equivalent form
\begin{equation}\label{eq:small-k1}
    \omega(k) = - 1 + k\, g^{-1}(k) - G(g^{-1}(k)) - \ln k\;.
\end{equation}

Now by using $\frac{d}{dz}\big(k\,z - G(z)\big) \vert_{z = g^{-1}(k)}
= 0\,$, we infer that $\omega$ has derivative
\begin{equation}
    \omega^\prime(k) = g^{-1}(k) - 1/k\;,
\end{equation}
or equivalently,
\begin{equation}
    \left(k^{-1}+\omega^\prime(k)\right)\Big\vert_{k = g(z)} = z\;.
\end{equation}
Comparison with (\ref{eq:Voiculescu}) then shows that $\omega^\prime
(k) = R(k)$ coincides with the Voiculescu $R$-transform. Hence, by
integrating,
\begin{equation}\label{eq:GM}
    \omega(k) = \int_0^k R(t)\, dt \;.
\end{equation}

The reasoning above makes good sense as long as $g(a) < k < g(b)$, so
that the saddle $z_0 = g^{-1}(k)$ lies outside the spectrum $[a,b]$.
It should be mentioned that the same result (\ref{eq:GM}) was
established under somewhat different assumptions (see below) by
rigorous analysis \cite{GM} using a large deviation principle.

\subsection{Asymptotics for $k$ large}\label{sect:large-k}

We turn to the complementary case of large $k$, meaning the ranges $k
< g(a) < 0$ and $k > g(b) > 0$. In this case, as we shall see, one
charge dissociates from the fluid in the interval $[a,b]$ in response
to the strong force exerted by the external field term $\mathrm{e}^{
Nkx}$. Consequently, the assumptions leading to the formula
(\ref{eq:int-small-k}) are no longer met and we need to proceed in a
different manner.

Our modified procedure is as follows. Abandoning the contour integral
formula (\ref{eq:CIR}), we insert the identities (\ref{eq:mrz-x1a}),
(\ref{eq:DH}) into the eigenvalue integral representation
(\ref{eq:mrz-x0}) for $\Omega$ and use $\mathrm{S}_N$ permutation
symmetry to single out one eigenvalue, say the first one $x_1 =:
x\,$. In this way we obtain the exact expression
\begin{eqnarray}
    \Omega(Nk \Pi) = C_{3,\,N}\, k^{-N+1} \int_\mathbb{R}
    \mathrm{e}^{Nkx - N\,V(x)}\,\pi_{N-1,\,N}(x) \,dx\;,
    \label{eq:mrz-x1} \\ \pi_{N-1,\,N}(x) = Z_N^{-1}
    \int_{\mathbb{R}^{N-1}} \prod_{2\leq i < j} (x_i - x_j)^2
    \prod_{2\leq l \leq N} (x-x_l)\,\mathrm{e}^{-N\,V(x_l)} \,
    dx_l \;. \label{eq:mrz-x2}
\end{eqnarray}
The function $\pi_{N-1,\,N}(x)$ is a polynomial of degree $N-1$ in
the variable $x\,$. By a classical result \cite{BDS,szego} it is
actually the \emph{orthogonal} polynomial of degree $N-1$ associated
with the weight function $\mathrm{e}^{-N\, V(x)}$. We find it
convenient to choose the normalization constants $C_{3,\,N}$ and
$Z_N$ in such a way that $\pi_{N-1,\,N}(x) = x^{N-1} + \ldots$ is
monic.

Once again, we will use saddle analysis and the method of steepest
descent to calculate the integral (\ref{eq:mrz-x1}) for large $N$. To
prepare this step, we observe that the orthogonal polynomial $\pi_{N
-1,\,N}$ has a big number $N-1$ of real zeroes concentrated in a
small neighborhood of the interval $[a,b]$. The large-$N$ asymptotics
of the high-order polynomial $\pi_{N-1,\,N}$ divides the $k$-axis
into different regions. In fact, when the absolute value of $k$ is
large, the term $\mathrm{e}^{Nkx}$ pushes the saddle of the
$x$-integral away from $[a,b]$ and thus into the region where
$\pi_{N-1,\,N}$ does not oscillate but varies monotonically. The
large-$N$ asymptotic analysis of (\ref{eq:mrz-x1}) then is rather
straightforward; see below. On the other hand, as $k$ decreases below
a critical value the saddle merges with the fluid $[a,b]$, where
$\pi_{N-1,\,N}$ oscillates rapidly. The integral representation
(\ref{eq:mrz-x1}) then does not give a direct view of the large-$N$
asymptotics. Fortunately, this case has already been dealt with in
Section \ref{sect:small-k} using the alternative representation by a
contour integral.

For definiteness, from here on let $k > g(b)$. (For $k < g(a)$ the
argument goes just the same.) The main contribution to the integral
(\ref{eq:mrz-x1}) then comes from large values of $x\,$, where the
polynomial (\ref{eq:mrz-x2}) behaves as
\begin{equation}
    \pi_{N-1,\,N}(x) \sim \mathrm{e}^{(N-1) \int_a^b
    \ln(x-y) \, d\nu(y)} \sim \mathrm{e}^{N G(x)} \;.
\end{equation}
By inserting this asymptotic expression into the integral
(\ref{eq:mrz-x1}) we obtain
\begin{equation}\label{eq:int-large-k}
    \omega(k) = \gamma_1 - \ln k + \lim_{N\to \infty} N^{-1}
    \ln \int \mathrm{e}^{N k x - N\,V(x) + N G(x)} d x
\end{equation}
with another constant $\gamma_1\,$.

The condition for $x \equiv x_0$ to be an extremum of the integrand
is
\begin{equation}
    k = V^\prime(x_0) - g(x_0)\;.
\end{equation}
{}From (\ref{eq:ana-cont}) we know that $V^\prime(z) - g(z) = h(z)$
is the second branch of the double-valued function $z \mapsto (g(z),
h(z))$. Thus the saddle-point equation is
\begin{equation}\label{eq:saddle}
    k = h(x_0) \;.
\end{equation}
This equation does not always have a solution for $x_0 \in \mathbb{R}
\,$. However, since $V$ is analytic and convex by assumption, the
derivative
\begin{equation}
    h^\prime(x) = \frac{d}{dx} \left( V^\prime(x) - g(x) \right) =
    V^{\prime\prime}(x) + \int_a^b \frac{d\nu(y)}{(x - y)^2}
\end{equation}
for $x > b$ is positive, so $h(x)$ is monotonically increasing.
Therefore if $k > h(b) = g(b)$, then $k$ lies in the range of the
function $h:\,(b, +\infty) \to \mathbb{R}\,$, and by the implicit
function theorem the monotonicity of $h$ guarantees the local
existence of an inverse
\begin{equation}
    k \mapsto h^{-1}(k) = z
\end{equation}
of the function $z \mapsto h(z) = k\,$. By $h^\prime(x) > 0$ for $x >
b $ the solution $x_0 = h^{-1}(k)$ of (\ref{eq:saddle}) is a maximum
of the integrand. Hence, by applying the method of steepest descent
and evaluating the integrand of (\ref{eq:mrz-x1}) at the critical
point $x_0 = h^{-1}(k)$ we obtain
\begin{eqnarray}
    \omega(k) = \gamma_1 + k\, h^{-1}(k) - V(h^{-1}(k)) +
    \int_a^b \ln \left( \frac{h^{-1}(k) - x}{k} \right)\, d\nu(x)
    \cr = \gamma_1 + \ell + k\, h^{-1}(k) - H(h^{-1}(k))
    - \ln k \;, \label{eq:large-k}
\end{eqnarray}
where the second equality is by (\ref{eq:mrz-i1}). We observe that
the derivative of $\omega$ is still given by the $R$-transform:
\begin{equation}
    \omega^\prime(k) = h^{-1}(k) - 1/k = R(k) \;.
\end{equation}
By matching (\ref{eq:large-k}) to the small-$k$ result
(\ref{eq:small-k}) we infer that $\gamma_1 + \ell = -1$. (Matching
will be justified in Section \ref{sect:match}.)

Essentially the same reasoning goes through for the opposite range
$k<g(a)=h(a)<0\,$, resulting in the same formula (\ref{eq:large-k}).
Notice that the argument of the logarithm under the integral sign in
(\ref{eq:large-k}) remains positive in this $k$-range, as there is a
sign change in both the numerator and the denominator.

\begin{figure}
    \begin{center}
        \epsfig{file=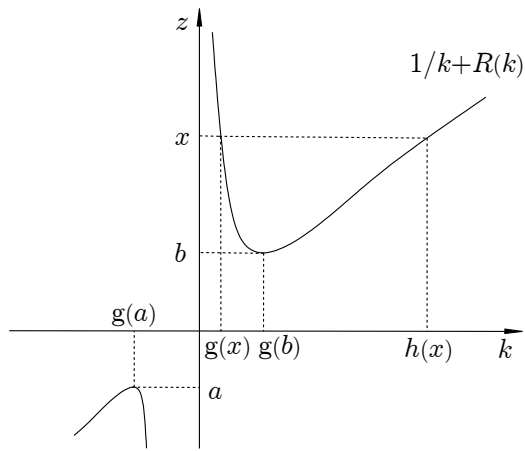,height=6cm}
    \caption{The saddle trajectories $k \mapsto g^{-1}(k)$ for
    $g(a) < k < g(b)$, and $k \mapsto h^{-1}(k)$ for $k < g(a)$
    or $g(b) < k$, piece together at the critical points $g(a)$,
    $g(b)$ to yield a smooth function $k \mapsto k^{-1} + R(k)$.}
    \label{fig:R-Transform}
    \end{center}
\end{figure}

To summarize our results, we have found that
\begin{equation}
    \hspace{-1cm}
    \omega(k) = \left\{ \begin{array}{ll} -1 + k\, g^{-1}(k)
    - G(g^{-1}(k)) - \ln k \;, &\quad g(a) < k < g(b)\;, \cr
    -1 + k\, h^{-1}(k) - H(h^{-1}(k)) - \ln k \;, &\quad
    k < g(a) ~ {\rm or} ~ g(b) < k \;. \end{array} \right.
\end{equation}
Recalling from Section \ref{sect:Coulomb} the relations $g(x) = h(x)$
and $G(x) = H(x)$ for $x \in \{a,b\}$, we see that the different
pieces of $\omega$ combine to a smooth function (Figure
\ref{fig:R-Transform}), since each piece has derivative
$\omega^\prime(k) = R(k)$.

\subsection{Absence of phase transition}\label{sect:match}

Related to the material of Sections \ref{sect:small-k} and
\ref{sect:large-k} there exist mathematical results \cite{GM} by
Guionnet and Maida (GM), which we now discuss briefly. Using large
deviations techniques, GM control the large-$N$ asymptotics of the
spherical integral (\ref{eq:HCIZ}) under the condition that the
empirical measure (i.e., the sum of $\delta$-functions located at the
eigenvalues) of a sequence of diagonal $N \times N$ matrices $X_N$
converges weakly ($N \to \infty$) to a compactly supported density of
states $\nu$ with $\mathrm{supp}\, \nu = [a,b]$. In contrast to the
present setting, GM stipulate that all eigenvalues of $X_N$ remain
strictly confined to the interval $[a,b]$. In Coulomb gas language
this means that infinitely high potential walls are placed at the
boundaries of the interval $[a,b]$.

With these assumptions, GM prove that $N^{-1} \ln I_{X_N}(k)$
converges to our result $\omega(k) = \int_0^k R(t) dt$ in the
small-$k$ range $g(a) < k < g(b)$. For $k$ outside this range,
however, they establish a qualitatively different behavior, separated
from the small-$k$ behavior by phase transitions at the critical
points $k = g(a)$ and $k = g(b)$. Please be assured that there is no
contradiction with our result (\ref{eq:large-k}), as the underlying
model assumptions are different. As we have seen, for large values of
$k$ the strong external field term $\mathrm{e}^{N k x}$ has the
effect of expelling one charge from the fluid of the other charges.
Such a dissociation phenomenon is forbidden by the assumptions of GM.

Still, in view of the phase transitions found by GM and the rather
different integral representations (\ref{eq:int-small-k}) and
(\ref{eq:int-large-k}) appearing in our treatment, the reader may
wonder why it is that the large-$N$ asymptotics for the small-$k$ and
large-$k$ regions match perfectly to give a smooth result for
$\omega(k)$. Let us now give a simple argument why this \emph{must}
be the case under our assumptions on the confining potential $V$.

Consider the mapping $\pi : \, \mathrm{End}(\mathbb{C}^N) \to
\mathbb{R}$ which sends $H$ to its projection $\pi(H) =
x$ to the one-dimensional subspace with projector $\Pi\,$. Imagine
computing the push forward of $\mu_N$ by this mapping,
\begin{equation}
    \mathrm{e}^{-N\,\mathrm{Tr}\,V(H)} dH\; \stackrel{\pi}{\mapsto}
    \; \mathrm{e}^{- N\,\Gamma(x) + O(N^0)} dx \;,
\end{equation}
for large $N$. In the language of quantum field theory one would call
$\Gamma$ the (large-$N$ limit of the) generating function for the
one-particle irreducible vertex functions.

Since $V$ is uniformly convex (i.e., its second derivative is bounded
below by a positive constant) and analytic, so is $\Gamma$. Indeed,
the latter results from the former as the fixed point of a
renormalization-group type recursive process of integrating out
variables. Now,
\begin{eqnarray}
    \omega(k) = \lim_{N\to\infty} N^{-1} \ln \int \mathrm{e}^{Nk\,
    \mathrm{Tr}\; \Pi\, H} d\mu_N(H) \cr = \lim_{N \to \infty} N^{-1}
    \ln \int \mathrm{e}^{Nkx -N\,\Gamma(x)} dx = \big( kx - \Gamma(x)
    \big) \Big\vert_{x : \, \Gamma^\prime(x) = k} \;.
\end{eqnarray}
Thus our function $\omega$ is the Legendre transform of $\Gamma$.
Since $\Gamma(x)$ is analytic and uniformly convex as a function of
$x \in \mathbb{R}$, the Legendre transform $\omega(k)$ must be
analytic as a function of $k \in \mathbb{R}$, ruling out any
possibility for a phase transition to occur.

Moreover, under the stated conditions on $V$ one can show that the
$R$-transform $R(k)$ is an entire function of $k \in \mathbb{C}$.
Also, in the present setting there is no reason to think that
$\omega$ should have any singularities in the complex $k$-plane.
Hence, by the identity principle (i.e., uniqueness of analytic
continuation) we expect that
\begin{equation}\label{eq:3.26}
    \omega(k) = \int_0^k R(t)\,dt
\end{equation}
holds for all $k \in \mathbb{C}$. This is the result we propose.

\section{Fermion-fermion sector}
\label{sect:FF}

Our argument so far deals with the rank-one case $K = \varphi \otimes
\widetilde{\varphi} = k \Pi\,$. It extends to the higher-rank case of
several replicas $\varphi_1\,, \ldots, \varphi_p$ without difficulty.
However, as we have seen in Section \ref{sect:motiv}, the
supersymmetry method requires as input the characteristic function
$\Omega(K)$ for $K = \varphi \otimes \widetilde{\varphi} + \psi
\otimes \widetilde{\psi}$, which includes the fermion-fermion
expression $\psi \otimes \widetilde{\psi}$ built from $\psi \in
\mathbb{C}^N \otimes \wedge((\mathbb{C}^N)^\ast)$ and
$\widetilde{\psi} \in (\mathbb{C}^N )^\ast \otimes
\wedge(\mathbb{C}^N)$. The reasoning of Section \ref{sect:Dyson} does
not apply to that second summand. Therefore we now develop another
scheme, concentrating again on the simple case of a single replica,
$K = \psi \otimes \widetilde{\psi}$.

So, in this section we make a study of $\Omega(K) = \int
\mathrm{e}^{\mathrm{Tr} \,HK} d\mu_N (H)$ for $K = \psi \otimes
\widetilde{\psi}$. Since the probability measure $\mu_N$ is invariant
under conjugation $H \mapsto g H g^{-1}$ for $g \in \mathrm{U}_N\,$,
the characteristic function $\Omega(\psi \otimes\widetilde{\psi})$
has the property of depending only on the $\mathrm {GL}_N $-invariant
$\langle \widetilde{\psi}, \psi \rangle$. Although this property is
not explicit from the definition of $\Omega\,$, it can be made so by
averaging \emph{ad hoc} over $\mathrm{U}_N$-orbits. Hence we compute
\begin{equation}\label{eq:IN-FF}
    I_N(\psi \otimes \widetilde{\psi}\,;H):= \int_{\mathrm{U}_N}
    \mathrm{e}^{\mathrm{Tr}\; (\psi \otimes \widetilde{\psi})\,g^{-1} H
    g} dg = \int_{\mathrm{U}_N} \mathrm{e}^{- \langle \widetilde{\psi}
    g^{-1} , \, H g \psi \rangle} dg
\end{equation}
as a useful preparation for the subsequent process of integrating
against $d\mu_N(H)$.

We claim that the integral (\ref{eq:IN-FF}) has the following
alternative expression:
\begin{equation}\label{eq:formula-FF}
    I_N(\psi \otimes \widetilde{\psi}\,; H) = N!^{-1}
    \int_0^\infty \mathrm{Det} \big( t - \langle \widetilde{\psi},
    \psi \rangle \, H \big)\, \mathrm{e}^{-t} dt \;.
\end{equation}
This formula is proved in the next subsection. Having established it,
we will determine the large-$N$ asymptotics of $N^{-1} \ln \Omega(N
\psi \otimes \widetilde{\psi})$ in Section \ref{sect:4.2}.

\subsection{Proof of formula (\ref{eq:formula-FF})}

After expanding the integrand of (\ref{eq:IN-FF}) by the power series
of the exponential function,
\begin{equation}
    \mathrm{e}^{- \langle \widetilde{\psi} g^{-1} , \, H g \psi \rangle}
    = \sum_{m=0}^N \frac{(-1)^m}{m!} \langle \widetilde{\psi} g^{-1} , \,
    H g \psi \rangle^m \;,
\end{equation}
our task is to compute the integrals $\int_{\mathrm{U}_N} \langle
\widetilde {\psi}g^{-1},\, H g \psi \rangle^m dg$. To do so, it is
helpful to recall some facts from the representation theory of
$\mathrm{GL}_N\,$.

The irreducible representations of $\mathrm{GL}_N$ are labeled by
highest weights $\lambda$ which are sequences of integers $\lambda
\equiv (\lambda_1, \lambda_2, \ldots, \lambda_N)$ with $\lambda_1
\geq \lambda_2 \geq \ldots \geq \lambda_N\,$. Here we will need only
those representations which extend holomorphically to
$\mathrm{End}(\mathbb{C}^N)$; these are distinguished by $\lambda_N
\geq 0$. The character of the representation $\rho_\lambda$ with
highest weight $\lambda$ is denoted by $s_\lambda(g) = \mathrm{Tr}\,
\rho_\lambda(g)$ and is called a Schur function. The Schur functions
form an orthonormal system w.r.t.\ Haar measure $dg$ for
$\mathrm{U}_N$ of total mass $\int_{\mathrm{U}_N} dg = 1:$
\begin{equation}
    \int_{\mathrm{U}_N} s_\lambda(g^{-1}) s_{\lambda^\prime}(g)
    \, dg = \delta_{\lambda \lambda^\prime} \;.
\end{equation}

To do the integrals $\int_{\mathrm{U}_N} \langle \widetilde {\psi}
g^{-1},\, H g \psi \rangle^m dg\,$, our first step is to recognize
the integrand as a special Schur function:
\begin{equation}\label{eq:schur1}
    \big( - \langle \widetilde{\psi} g^{-1} , \, H g \psi \rangle
    \big)^m = s_{[1^m]}((\psi \otimes \widetilde{\psi}) g^{-1} H g)\;,
\end{equation}
where $\lambda = [1^m]$ is the highest weight given by
\begin{equation}
    \lambda_1 = \lambda_2 = \ldots = \lambda_m = 1 \;, \quad
    \lambda_{m+1} = \ldots = \lambda_N = 0 \;.
\end{equation}
In the second step we use the general identity
\begin{eqnarray}
    \int_{\mathrm{U}_N} s_\lambda(K g^{-1} H g) \, dg
    = \int_{\mathrm{U}_N} \mathrm{Tr}\, \rho_\lambda(K)
    \rho_\lambda(g^{-1}) \rho_\lambda(H) \rho_\lambda(g)\, dg \cr
    = \frac{\mathrm{Tr}\, \rho_\lambda(K) \, \mathrm{Tr}\,
    \rho_\lambda(H)}{ \mathrm{Tr}\,\rho_\lambda(\mathrm{Id}_N)}
    = \frac{s_\lambda(K) \, s_\lambda(H)}{s_\lambda(\mathrm{Id}_N)} \;,
    \label{eq:schur2}
\end{eqnarray}
which derives from the fact that $\rho_\lambda$ is a representation.

We now establish (\ref{eq:schur1}). For this we start from the
relation
\begin{equation}\label{eq:cauchy}
    \mathrm{Det}\,(\mathrm{Id}_N - K) =
    \sum\nolimits_{m=0}^N (-1)^m s_{[1^m]}(K)\;,
\end{equation}
which is a special case of what is sometimes called the dual Cauchy
identity \cite{bump}. Then, substituting $K = \psi \otimes
\widetilde{\psi}$ we manipulate the left-hand side as follows:
\begin{eqnarray}
    \mathrm{Det}\,(\mathrm{Id}_N - \psi \otimes \widetilde{\psi})
    = \mathrm{e}^{\mathrm{Tr}\, \ln\, (\mathrm{Id}_N - \psi \otimes
    \widetilde{\psi})} = \mathrm{e}^{- \sum_{m=1}^N m^{-1} \mathrm{Tr}
    \, (\psi \otimes \widetilde{\psi})^m}\cr
    = \mathrm{e}^{+ \sum_{m=1}^N
    m^{-1} \langle \widetilde{\psi}, \psi \rangle^m}
    = \mathrm{e}^{- \ln\, (1 - \langle \widetilde{\psi},\psi \rangle)}
    = \sum\nolimits_{m=0}^N \langle \widetilde{\psi},\psi \rangle^m \;.
\end{eqnarray}
By combining equations we have
\begin{equation}
    \sum\nolimits_{m=0}^N \langle \widetilde{\psi},\psi \rangle^m =
    \sum\nolimits_{m=0}^N (-1)^m s_{[1^m]}(\psi\otimes\widetilde{\psi})\;.
\end{equation}
Since the Schur function $s_{[1^m]}(K)$ is a homogeneous polynomial
of degree $m$ in the matrix entries of $K$, the desired relation
(\ref{eq:schur1}) follows on replacing $\widetilde{\psi} \to
\widetilde{\psi} g^{-1} H g$.

We are now in a position to compute our integral:
\begin{eqnarray}
    \int_{\mathrm{U}_N} \big( - \langle \widetilde {\psi} g^{-1},
    H g \psi \rangle \big)^m dg \cr = \int_{\mathrm{U}_N} s_{[1^m]}
    (\psi \otimes \widetilde {\psi} g^{-1} H g) dg =
    \frac{s_{[1^m]}(\psi \otimes \widetilde{\psi})
    s_{[1^m]}(H)}{s_{[1^m]}(\mathrm{Id}_N)}\;,
\end{eqnarray}
where we used (\ref{eq:schur2}). On the right-hand side we have the
simplification $s_{[1^m]}(\psi \otimes \widetilde{\psi}) = (-\langle
\widetilde{\psi} , \psi \rangle)^m$ from (\ref{eq:schur1}), and the
denominator is the dimension of the representation:
\begin{equation}
    s_{[1^m]}(\mathrm{Id}_N) = \mathrm{dim}\, \wedge^m (\mathbb{C}^N)
    = \frac{N!}{m!(N-m)!} \;.
\end{equation}
Thus we obtain the following expression for $I_N:$
\begin{eqnarray}
    I_N(\psi\otimes\widetilde{\psi}\,;H) = \sum_{m=0}^N \frac{1}
    {m!} \int_{\mathrm{U}_N} \big( - \langle \widetilde {\psi}g^{-1},H
    g\psi \rangle\big)^m dg \cr = \frac{1}{N!} \sum_{m=0}^N (-1)^m
    (N-m)! \langle \widetilde{\psi}, \psi \rangle^m s_{[1^m]}(H)\;.
\end{eqnarray}

We finally resum this expansion. For that, we express the factorial
as $(N-m)! = \int_0^\infty t^{N-m} \mathrm{e}^{-t} dt$ and use the
dual Cauchy identity (\ref{eq:cauchy}) in reverse to arrive at
\begin{eqnarray}
    I_N(\psi\otimes\widetilde{\psi}\,;H) = \int_0^\infty
    \frac{t^N}{N!} \sum_{m=0}^N (-1/t)^m \langle \widetilde{\psi},
    \psi \rangle^m s_{[1^m]}(H)\; \mathrm{e}^{-t} dt \cr = N!^{-1}
    \int_0^\infty \mathrm{Det} \big(t - \langle \widetilde{\psi},
    \psi \rangle \, H \big) \, \mathrm{e}^{-t} dt \;.
\end{eqnarray}
This completes the proof.

\subsection{Large-$N$ limit}\label{sect:4.2}

Recall the definition
\begin{equation}
    \Omega(N \psi \otimes \widetilde{\psi}) = \int I_N (N \psi
    \otimes \widetilde{\psi}\,;H)\, d\mu_N(H)\;.
\end{equation}
As was reviewed in Section \ref{sect:motiv}, in the superbosonization
method one lifts $\Omega(N\psi \otimes \widetilde{\psi})$ to a
holomorphic function $k \mapsto \widehat{\Omega}(Nk)$ of a complex
variable $k \in \mathbb{C}\,$. By substituting $t \to Nt$ in
(\ref{eq:formula-FF}) and changing the order of integration, we see
that
\begin{equation}
    \widehat{\Omega}(Nk) = \frac{N^{N+1}}{N!} \int_0^\infty \left( \int
    \mathrm{Det}\,( t - k H)\, d\mu_N(H) \right) \mathrm{e}^{-Nt} dt
\end{equation}
is such a function. We now tackle the task of determining its
large-$N$ limit.

Stirling's formula gives $\lim_{N\to\infty} N^{-1} \ln\,(N^{N+1} /
N!) = 1$, and by the type of reasoning of Section \ref{sect:Dyson} we
find the large-$N$ approximation
\begin{equation}
    \int \mathrm{Det}\,( t - k H)\, d\mu_N(H) \sim
    \mathrm{e}^{N \int_a^b \ln\,(t-kx)\, d\nu(x)} \;.
\end{equation}
We now take $N \to \infty$ to define
\begin{equation}
    \phi(k) := \lim_{N\to\infty} N^{-1} \ln \widehat{\Omega}(Nk) \;.
\end{equation}
Note that $\phi(0) = 0$, and by combining the above we have
\begin{equation}
    \phi(k) = 1 + \lim_{N\to\infty} N^{-1} \ln \int_0^\infty
    \mathrm{e}^{-Nt + N \int_a^b \ln\,(t-kx)\, d\nu(x)} dt \;.
\end{equation}

The integral on the right-hand side is a close cousin of the integral
(\ref{eq:int-small-k}) and, in fact, has essentially the same saddle
point, $t = k\,g^{-1}(k)$. The overall sign in the exponent is
reversed, but at the same time we are now integrating over $t$ in the
direction of the real axis, while the path of steepest descent in the
case of (\ref{eq:int-small-k}) was along the imaginary direction.
Thus the saddle analysis is essentially the same and need not be
repeated here. We just quote the result:
\begin{equation}
    \phi(k) = 1 - k\,g^{-1}(k) + \int_a^b
    \ln\,(k\, g^{-1}(k) - k x)\, d\nu(x) \;.
\end{equation}
This is exactly the negative of $\omega(k)$ in (\ref{eq:small-k}).
Hence we conclude (cf.\ (\ref{eq:3.26})) that
\begin{equation}
    \phi(k) = - \int_0^k R(t)\, dt \;.
\end{equation}
Returning to the original meaning of $k = \langle \widetilde{\psi} ,
\psi \rangle$ we state our final result as follows:
\begin{equation}
    \lim_{N \to \infty} N^{-1} \ln\, \Omega(N \psi \otimes
    \widetilde{\psi}) = - \sum_{n\geq 1} \frac{c_n}{n}\,
    \langle \widetilde{\psi} , \psi \rangle^n \;.
\end{equation}

In Section \ref{sect:Dyson} we achieved a satisfactory understanding
of the large-$N$ behavior of $N^{-1} \ln \Omega(NK)$ for a rank-one
operator $K = \varphi \otimes \widetilde{\varphi}$. The current
section does the same for $K = \psi \otimes \widetilde{\psi}$. In
both cases, our treatment carries over without difficulty to $K$ of
higher rank (or several replicas).

However, the supersymmetry method calls for $\Omega(NK)$ with $K$ of
the \emph{mixed} type $K = \varphi \widetilde{\varphi} + \psi
\widetilde{ \psi}\,$; cf.\ (\ref{eq:int-formula}). The techniques
used in Sections \ref{sect:Dyson} and \ref{sect:FF} are quite
different, and at the present time we do not know how to combine them
to handle the mixed situation. For this reason we now turn to a
completely different approach.

\section{Large-$N$ combinatorial theory of $\Omega(K)$}
\label{sect:combinatorics}

In the present section we launch yet another attack, combining the
planar graph limit of perturbation theory with Speicher's
combinatorial approach to the free cumulants. An early physics paper
in this direction is \cite{zee}.

\subsection{From moments to cumulants}

We begin by reviewing the usual connection between the moments and
the cumulants of a random number, and afterwards state the analog of
this connection in the setting of free probability theory.

\subsubsection{Commutative case ($N = 1$).}

Although our interest is in $N \times N$ random matrices in the limit
$N \to \infty\,$, we temporarily set $N = 1$ and review the
connection between moments and cumulants for the case of a single
random variable $x \in \mathbb{R}$ with probability measure $\mu$.
The moments $m_n$ of $\mu$,
\begin{equation}
    m_n = \int_\mathbb{R} x^n d\mu(x) \;,
\end{equation}
are generated by the characteristic function,
\begin{equation}
    \Omega(k) = \int_\mathbb{R} \mathrm{e}^{kx} d\mu(x) =
    \sum_{n=0}^\infty m_n \frac{k^n}{n!}\;, \qquad \Omega(0) = 1\;,
\end{equation}
while the cumulants $c_n$ are generated by the logarithm of
$\Omega(k)$:
\begin{equation}
    \ln\, \Omega(k) = \sum_{n=1}^\infty c_n \frac{k^n}{n!} \;.
\end{equation}
Sometimes one wants to express the moments in terms of the cumulants,
or vice versa. Switching between the two descriptions is an exercise
in basic combinatorics and the Leibniz product rule of differential
calculus. Let us do this exercise as a warm up for a more strenuous
calculation to come in Section \ref{sect:speicher} below.

We start by writing the $n^\mathrm{th}$ moment as an $n^\mathrm{th}$
derivative at $k = 0:$
\begin{equation}\label{eq:mom-by-dif}
    m_n = \frac{d^n}{dk^n}\,\mathrm{e}^{\ln\Omega(k)}\Big\vert_{k=0}
    = \left( \mathrm{e}^{- \ln \Omega(k)} \frac{d}{dk} \circ
    \mathrm{e}^{\ln \Omega(k)} \right)^n \Bigg\vert_{k=0} \;.
\end{equation}
By the Leibniz rule this becomes
\begin{equation}
    m_n = \left( \frac{d}{dk} + \frac{\Omega^\prime(k)}{\Omega(k)}
    \right)^n \Bigg\vert_{k=0} \;.
\end{equation}
From this formula, by multiplying out factors using the distributive
law, we generate a sum of terms which are in one-to-one
correspondence with partitions $p \in \Pi(n)$ of the set $\{ 1, 2,
\ldots, n \}$. Indeed, we are to break up (or partition) $n$
identical factors into $\nu_l(p)$ blocks of integer length $l \geq 1$
subject to $\sum_l l\, \nu_l (p) = n\,$. Each block of length $l$
consists of one logarithmic derivative $\Omega^\prime (k) /
\Omega(k)$ together with $l-1$ derivatives $d/dk$ acting on it before
evaluation at $k = 0\,$. Now since each block of length $l$
contributes a cumulant $c_l = (d / dk)^l \ln \Omega(k) \vert_{k=0}\,
$, we obtain a formula expressing the moment,
\begin{equation}\label{eq:mom-cum}
    m_n = \sum_{p \in \Pi(n)} \; \prod_{l \ge 1} c_l^{\nu_l(p)} \;,
\end{equation}
as a sum over partitions $p \in \Pi(n)$ in terms of the cumulants
$c_1, \ldots, c_n\,$.

\subsubsection{Speicher's formula ($N = \infty$).}

We return to the case of $N \times N$ random matrices $H$ with
probability measure $\mu_N$ and recall the definition
(\ref{eq:moments}) of the moments $m_{n,\,N}$ and their large-$N$
limits $m_n = \lim_{N\to\infty} m_{n,\,N}\,$. We also recall the
definition (\ref{eq:Voiculescu}) of the free cumulants $c_n$ by the
Voiculescu $R$-transform.

It turns out that the large-$N$ moments $m_n$ are expressible in
terms of the free cumulants $c_l$ (with $l \leq n$) by a
combinatorial formula closely analogous to (\ref{eq:mom-cum}). Due to
Speicher \cite{speicher}, this formula reads
\begin{equation}\label{eq:speicher}
    m_n = \sum_{p\in\mathrm{NC}(n)}\;\prod_{l\ge 1} c_l^{\nu_l(p)}\;.
\end{equation}
It differs from (\ref{eq:mom-cum}) only in that the sum over all
partitions $p \in \Pi(n)$ has been restricted to the subset of
\emph{non-crossing} partitions $p \in \mathrm{NC}(n)$. A partition
$p$ is called non-crossing if for any two pairs $\{ i_1, i_2 \}$ and
$\{ j_1 , j_2 \}$ taken from any two different blocks of $p$, it
never happens that
\begin{equation}
    i_1 < j_1 < i_2 < j_2 \;.
\end{equation}
Informally speaking, this means that if we arrange the numbers $1,
\ldots, n$ in cyclic order around the boundary of a disk and connect
the numbers of each block of the partition by lines via the interior
of the disk in a `minimal' way, then the lines associated with
different blocks do not cross each other. See Figure \ref{fig:bild1}
below for an example.

Consider the single-block (or trivial) partition $p$, where $\nu_1(p)
= \ldots = \nu_{n-1}(p) = 0$ and $\nu_n(p) = 1$. This partition is
non-crossing and contributes $\prod_{l\ge 1} c_l^{\nu_l(p)} = c_n$ to
Speicher's formula (\ref{eq:speicher}) for the moment $m_n\,$. Thus
(\ref{eq:speicher}) has the general structure
\begin{equation}
    m_n = c_n + P_{n-1}(c_1,c_2,\ldots,c_{n-1}) \;,
\end{equation}
where $P_{n-1}$ is a polynomial in the free cumulants of order $n-1$
or less. It follows that the relation (\ref{eq:speicher}) can be
inverted to express $c_n$ in terms of the moments $m_1, \ldots,
m_n\,$. In this sense the formula (\ref{eq:speicher}) may serve to
define the free cumulants in terms of the large-$N$ moments. Speicher
\cite{speicher} has proved (under suitable conditions on $\mu_N$)
that this combinatorial definition is equivalent to Voiculescu's
analytic definition of $k \mapsto R(k) = \sum c_n\, k^{n-1}$ via
inversion of $z \mapsto g(z) = \sum m_n \, z^{-n-1}$.

\subsection{Cumulant tensor}

After these preliminaries, we will outline a combinatorial
description of the large-$N$ limit of $N^{-1} \ln \Omega(N K)$ by way
of expansion in powers of $K$ (restricting the numerical part of $K$
to be of finite rank). For this purpose we fix some degree $n \geq 1$
and consider the full tensor of cumulants
\begin{equation}\label{eq:nth-cum}
    C_{j_1 \ldots j_n}^{\, i_1 \ldots \, i_n} = \frac{\partial^n}
    {\partial K_{i_1 j_1} \partial K_{i_2 j_2} \cdots
    \partial K_{i_n j_n}} \ln \, \Omega(K) \Big\vert_{K = 0} \;.
\end{equation}
We refer to the set of components $C_{j_1 \ldots j_n}^{\, i_1 \ldots
\, i_n}$ as the `cumulant tensor' (at degree $n$) for short. Please
note that our simple notation does not display the dependency which
$C_{j_1 \ldots j_n}^{\, i_1 \ldots \, i_n}$ has on the random matrix
dimension $N$.

For a general probability measure $\mu_N$ with no symmetries (and $N
\geq n$), we expect all components of this tensor to be independent
of each other. However, we shall now assume a probability
distribution of the $\mathrm{U}_N$-invariant form $\mathrm{e}^{-N\,
\mathrm{Tr} \, V(H)} dH$. The characteristic function then inherits
the invariance property $\Omega(K) = \Omega(g^{-1} K g)$ for $g \in
\mathrm{U}_N\,$. It follows that the set of components $C_{j_1 \ldots
j_n}^{\, i_1 \ldots \, i_n}$ constitute an invariant tensor of
$\mathrm{U}_N$ (or by analytic extension, of $\mathrm{GL}_N \equiv
\mathrm{GL}_N (\mathbb{C})$, the complexification of $\mathrm{U}_N$):
\begin{equation}
    C_{j_1 \ldots j_n}^{\,i_1 \ldots\, i_n} = \sum_{i_1^\prime,\ldots,
    i_n^\prime, j_1^\prime, \ldots, j_n^\prime} g^{i_1}_{i_1^\prime}
    \cdots g^{i_n}_{i_n^\prime} \; C_{j_1^\prime \ldots j_n^\prime}^{\,
    i_1^\prime \ldots \, i_n^\prime}\; (g^{-1})^{j_1^\prime}_{j_1}
    \cdots (g^{-1})^{j_n^\prime}_{j_n} \;,
\end{equation}
$g \in \mathrm{GL}_N \,$.
By a classical result of invariant theory due to Weyl \cite{weyl},
one then knows that $C_{j_1 \ldots j_n}^{\,i_1 \ldots\, i_n}$ is
nonzero only if the numbers $\{ j_1 , \ldots, j_n \}$ agree as a set
with the set $\{ i_1 , \ldots, i_n \}$, i.e., if there exists an
element $\pi$ of the symmetric group $\mathrm{S}_n$ such that $j_l =
i_{\pi (l)}$ for $l = 1, \ldots, n\,$. Thus our cumulant tensor can
be expressed as a sum over permutations:
\begin{equation}\label{eq:cum-invs}
    C_{j_1 \ldots j_n}^{\, i_1 \ldots \, i_n} = \sum_{\pi \in
    \mathrm{S}_n} \gamma_{n,\,N}(\pi) \prod_{l=1}^n \delta_{i_l ,
    \, j_{\pi(l)}} \;.
\end{equation}
Moreover, by its definition (\ref{eq:nth-cum}) as an $n^\mathrm{th}$
symmetric derivative, $C_{j_1 \ldots j_n}^{\, i_1 \ldots \, i_n}$ is
invariant w.r.t.\ any permutation $\sigma \in \mathrm{S}_n$ of the
index pairs, $(i_l , j_l) \mapsto (i_{\sigma (l)}, j_{\sigma (l)} )$,
$l = 1, \ldots, n\,$. It follows that we may assume the coefficients
$\gamma_{n,\,N}(\pi)$ to be conjugacy class functions, i.e.,
$\gamma_{n,\,N} (\pi) = \gamma_{n,\,N} (\sigma^{-1} \pi \sigma)$ for
all $\sigma \in \mathrm{S}_n\,$.

Recall now from the basic theory of the symmetric group that the
conjugacy class $[\pi]$ of an element $\pi \in \mathrm{S}_n$ is
determined by its cycle structure; more precisely, by the set of
non-negative integers $\nu_1(\pi), \nu_2(\pi), \ldots, \nu_n(\pi)$
subject to $\sum_l l \nu_l(\pi) = n$ where $\nu_l(\pi)$ is the number
of cycles of $\pi$ of length $l$. An important role in the following
will be played by the conjugacy class, $[\mathrm{irr}]$, of all
irreducible cycles -- by this we mean the conjugacy class of elements
$\pi \in \mathrm{S}_n$ with $\nu_1(\pi) = \ldots = \nu_{n-1}(\pi) =
0$ and $\nu_n(\pi) = 1$.

\subsection{Large-$N$ hypothesis}

What has been said so far is true for all $N$, but now we take the
large-$N$ limit and claim the following \emph{large-$N$ hypothesis}:
\begin{equation}\label{hypothesis}
    \lim_{N \to \infty} N^{n-1}\gamma_{n,\,N}(\pi) = \left\{
    \begin{array}{ll} c_n &{\rm if}\; [\pi] = [\mathrm{irr}]\;,
    \cr 0 &{\rm else}. \end{array} \right.
\end{equation}
In words: $N^{n-1}\gamma_{n,\,N}(\pi)$ goes to zero for $N\to \infty$
unless $\pi \in \mathrm{S}_n$ belongs to the conjugacy class of one
irreducible cycle of length $n\,$. (If so, the cumulant tensor of
degree $n$ is determined by a single number $c_n$ for $N\to\infty\,$.
We shall see that $c_n$ is in fact the $n^\mathrm{th}$ free cumulant,
vindicating our notation.) The hypothesis (\ref{hypothesis}) is
present in \cite{collins}, Theorem 4.5. To make our paper
self-contained, we now offer some motivation (if not rigorous
justification) of this hypothesis from perturbation theory.

\subsection{Perturbation theory argument}

We here assume the exponent of our probability distribution
$\mathrm{e}^{-N\,\mathrm{Tr}\,V(H)} dH$ to be of the form
\begin{equation}
    \mathrm{Tr}\, V(H) = \frac{1}{2\sigma^2} \mathrm{Tr}\, H^2 +
    \mathrm{Tr}\, W(H) \;,
\end{equation}
where the parameter $\sigma^2$ is small, so that the interaction (or
non-quadratic part) $\mathrm{Tr}\, W(H)$ can be treated as a
perturbation. In order to develop a perturbation expansion in the
small parameter $\sigma^2$, we single out the Gaussian (or GUE)
measure
\begin{equation}
    d\mu_\mathrm{GUE}(H) = \mathrm{e}^{-\frac{N}{2\sigma^2}
    \mathrm{Tr}\, H^2} dH \;,
\end{equation}
with Lebesgue measure $dH$ normalized by $\int d\mu_\mathrm{GUE}(H) =
1$. Our goal is to compute
\begin{equation}
    \Omega(NK) = \int \, \mathrm{e}^{-N\,\mathrm{Tr}\,
    W(H) + N\,\mathrm{Tr}\,HK}\, d\mu_\mathrm{GUE}(H)\;.
\end{equation}
We take it for granted that $\Omega(0) = \int \mathrm{e}^{-N\,
\mathrm{Tr}\, W(H)} d\mu_\mathrm{GUE}(H) = 1$ by the choice of
normalization constant $W(0)$.

Now, by passing to the logarithm on both sides of the equation and
shifting the integration variable $H \to H + \sigma^2 K$, we obtain
\begin{equation}
    \ln \Omega(N K) = \frac{N \sigma^2}{2} \mathrm{Tr}\, K^2 +
    \ln \int \, \mathrm{e}^{-N\,\mathrm{Tr}\, W(H + \sigma^2 K)}\,
    d\mu_\mathrm{GUE}(H) \;.
\end{equation}
Next we use the trick of writing the $\mathrm{GUE}$ measure as the
result of applying the heat semigroup (generated by the Laplacian
$\Delta = \sum \partial^2 / \partial H_{ij} \partial H_{j\,i}$) to
the Dirac $\delta$-distribution with unit mass localized at $H = 0:$
\begin{equation}
    d\mu_\mathrm{GUE}(H) =
    \big( \mathrm{e}^{\sigma^2 \Delta /2N}\delta \big)(H) \;.
\end{equation}
We then use partial integration to bring $N^{-1} \ln \Omega(NK)$ into
the form
\begin{equation}\label{eq:PT}
    N^{-1} \ln \Omega(N K) = \frac{\sigma^2}{2} \mathrm{Tr}\, K^2
    + N^{-1} \ln \left( \mathrm{e}^{\sigma^2 \Delta / 2N} \mathrm{e}^{
    -N\, \mathrm{Tr}\, W(H)} \right) \Big\vert_{H = \sigma^2 K}\;.
\end{equation}
This expression serves as our starting point to develop the
perturbation expansion as follows (we give only a brief sketch,
referring to the literature \cite{large-N} for greater detail).

One expands the exponential function $\mathrm{e}^{- N\, \mathrm{Tr}
\, W(H)}$ by its power series. Using standard graphical code, one
represents each monomial $\mathrm{Tr}\, H^l$ in $\mathrm{Tr}\, W(H)$
by an $l$-vertex. One also expands the heat operator $\mathrm{e}^{
\sigma^2 \Delta / 2N}$ and represents each action of the Laplacian
$\Delta$ by an edge. In this way the contributions to the r.h.s.\ of
(\ref{eq:PT}) are drawn as graphs.

Any given graph contributes with an overall power of $N^{\chi-1}$
where $\chi$ is the Euler characteristic of the graph, i.e., the
number of vertices minus the number of edges plus the number of
faces. Indeed, every vertex carries a factor of $N$, every edge comes
with a factor of $N^{-1}\sigma^2$, and every face corresponds to a
free summation variable, thereby contributing one factor of $N$. (To
do this counting, we first delete the vertex legs that are saturated
by the final substitution $H \to \sigma^2 K$.) Thus the second term
on the r.h.s.\ of (\ref{eq:PT}) can be organized as a sum over
topological sectors:
\begin{equation}\label{eq:PT1}
    N^{-1} \ln \left( \mathrm{e}^{\sigma^2 \Delta / 2N} \mathrm{e}^{
    -N\, \mathrm{Tr}\, W(H)} \right) \Big\vert_{H = \sigma^2 K} =
    \sum\nolimits_{\chi} N^{\chi-1} \omega_\chi(K) \;,
\end{equation}
where $\omega_\chi(K)$ is the sum of all contributions from graphs
with Euler characteristic $\chi$.

By the linked cluster theorem, disconnected graphs cancel upon taking
the logarithm. Now among the set of all connected graphs (with at
least one substitution $H \to \sigma^2 K$) the Euler characteristic
becomes maximal for planar graphs with the topology of a disk $D$,
where all substitutions $H \to \sigma^2 K$ are arranged around the
boundary of $D$. The Euler characteristic for such a graph is
$\chi(D) = 1$; indeed, a triangle, say, has three vertices, three
edges, and one face, so $\chi = 3 - 3 + 1 = 1$.

The Euler characteristic of a non-planar graph is known to be smaller
than the planar value $\chi(D) = 1$. There also exist planar graphs
which have a topology different from that of a disk; examples are the
annulus or a disk with several holes in its interior resulting in an
`inner' boundary. Again, the Euler characteristic of these non-disk
planar graphs is smaller than 1. For an annulus $A$ one has $\chi(A)
= \chi(D) - 1 = 0$, since there is one face missing as compared with
the disk $D$.

The upshot of all this is that the leading contribution to
(\ref{eq:PT1}) in the large-$N$ limit is of order $O(N^0)$ and is
given by the sum over all planar graphs with disk topology. Because
these graphs are constructed by inserting every substitution $H \to
\sigma^2 K$ into a single line (circulating around the boundary of
the disk-shaped graph), they all produce single-trace contributions
$\mathrm{Tr}\, K^n$, $n \geq 1$. Perturbation theory thus leads us to
expect
\begin{equation}\label{eq:expect}
    \lim_{N \to \infty} N^{-1} \ln\,\Omega(NK) = \omega_1(K)
    = \sum_{n=1}^\infty \frac{c_n}{n} \; \mathrm{Tr}\, K^n \;.
\end{equation}
The reason for writing the coefficient of $\mathrm{Tr}\, K^n$ as $c_n
/ n$ will become clear shortly.

It should be emphasized at this point that we are taking the limit $N
\to \infty$ while keeping the rank of (the numerical part of) $K$
\emph{finite}. Non-disk (e.g., annular) planar graphs make
contributions of the multi-trace form
\begin{equation}\label{multi-trace}
    N^{-r}\;\mathrm{Tr}\,(K^{n_1})\, \mathrm{Tr}\,(K^{n_2}) \cdots
    \mathrm{Tr}\,(K^{n_r})\,\mathrm{Tr}\,(K^{n-n_1-n_2-\ldots-n_r})\;,
\end{equation}
which would not become negligible in the large-$N$ limit if we let
$\mathrm{Tr} \, K^n$ grow with $N$.

This ends our excursion into perturbation theory. Based on the result
(\ref{eq:expect}), which should even be valid beyond the domain of
validity of perturbation theory (if only as a series expansion with
finite radius of convergence), it is straightforward to compute the
large-$N$ limit of the cumulant tensor (\ref{eq:nth-cum}). Doing so
with the help of (\ref{eq:cum-invs}), we immediately arrive at the
large-$N$ hypothesis (\ref{hypothesis}). Indeed, in the process of
applying the first derivative $\partial/ \partial K_{i_1 j_{\pi(1)}}$
to $\mathrm{Tr}\, K^n / n$ we have $n$ factors of $K$ to choose from
and thus a freedom which cancels the factor $1/n\,$; application of
the remaining $n-1$ derivatives $\partial/\partial K_{i_l j_{\pi(l)
}}$ gives precisely the sum over irreducible permutations $\sum_{\pi
\in [\mathrm{irr}]} \prod_l \delta_{i_l ,\, j_{\pi(l)}}$.

At the present stage, we cannot tell whether $c_n$ indeed is the
$n^\mathrm{th}$ free cumulant, but this open question will be settled
in the next subsection. Our strategy will be to show that the
large-$N$ hypothesis (\ref{hypothesis}) implies Speicher's formula
(\ref{eq:speicher}). Since it is already known that the coefficients
$c_n$ of the latter formula are the free cumulants, the desired
result follows.

\subsection{Retrieving Speicher's formula}
\label{sect:speicher}

Just as in the case (\ref{eq:mom-by-dif}) of a random number, we can
generate the moments (\ref{eq:moments}) of the probability measure
$\mu_N$ by differentiation of the characteristic function $\Omega(K)
= \mathrm{e}^{\ln \Omega(K)} = \int \mathrm{e}^{ \mathrm{Tr}\, HK}
d\mu_N(H)$ at $K = 0:$
\begin{equation}\label{eq:mom-dif}
    m_{n,\,N} = N^{-1} \sum_{i_1, \ldots, i_n} \frac{\partial^n}
    {\partial K_{i_n i_{n-1}} \cdots \partial K_{i_2 i_1} \partial
    K_{i_1 i_n}} \, \mathrm{e}^{\ln \Omega(K)} \Big\vert_{K = 0}\;,
\end{equation}
where indices are arranged in cyclic order and summed in order to
manufacture the expectation of a single trace $\mathrm{Tr}\,H^n$. By
the identity
\begin{equation}
    \mathrm{e}^{-\ln\Omega}\frac{\partial}{\partial K_{i_k i_{k-1}}}
    \circ \mathrm{e}^{\ln \Omega} = \frac{\partial} {\partial K_{i_k
    i_{k-1}}} + \frac{\partial \ln \Omega} {\partial K_{i_k i_{k-1}}}
\end{equation}
the expression for $m_{n,\,N}$ is cast in the form
\begin{equation}\label{eq:mom-f-cum}
    \hspace{-1cm}
    m_{n,\,N} = N^{-1} \sum_{i_1, \ldots, i_n} \left( \frac{\partial}
    {\partial K_{i_n i_{n-1}}} + \frac{\partial \ln \Omega} {\partial
    K_{i_n i_{n-1}}} \right) \cdots \left( \frac{\partial} {\partial
    K_{i_1 i_n}} + \frac{\partial \ln \Omega} {\partial K_{i_1 i_n}}
    \right) \Bigg\vert_{K = 0} .
\end{equation}
In the following we adopt the convention of assigning to $\partial /
\partial K_{i_k i_{k-1}}$ the number $k\,$.

Now, by evaluating the derivatives at $K = 0$ we get an expression
which is a polynomial in the cumulant tensor of (\ref{eq:nth-cum}).
The summands of the polynomial generated in this way are in
one-to-one correspondence (for fixed indices $i_1, \ldots, i_n$ and
by the bijection $\partial / \partial K_{i_k i_{k-1}} \leftrightarrow
k$) with partitions $p \in \Pi(n)$, where each block of length $l$ of
$p$ corresponds to a cumulant tensor of degree $l$. For example, the
following partition $p\in\mathrm{NC}(8) \subset \Pi(8)$,
\begin{equation}
    p = \{ 1, 5, 8\} \cup \{ 2, 3, 4 \} \cup \{ 6, 7 \}
\end{equation}
contributes to $m_{8,\,N}$ as
\begin{equation}\label{eq:example}
    N^{-1} \sum_{i_1, \ldots,\, i_8} C_{i_8 i_4 i_7}^{i_1 i_5 i_8}
    \, C_{i_1 i_2 i_3}^{i_2 i_3 i_4}\, C_{i_5 i_6}^{i_6 i_7} \;.
\end{equation}

\subsubsection{Contribution from non-crossing partitions.}

As a first step (which will turn out to be the main step), we compute
the contribution to (\ref{eq:mom-f-cum}) from the non-crossing
partitions $\mathrm{NC}(n)$, using the relations (\ref{eq:nth-cum}),
(\ref{eq:cum-invs}) and the large-$N$ hypothesis (\ref{hypothesis}).

So, let $p \in \mathrm{NC}(n)$. To prepare for the task of counting
powers of $N$, we will associate with $p$ a $2$-complex $\Sigma(p)$
as follows. Let $D$ be (any) disk, and divide the boundary line of
$D$ into $n$ segments numbered in counterclockwise order by $1,
\ldots, n\,$. These segments shall be 1-cells of the 2-complex
$\Sigma (p)$ to be constructed. Each pair $(k,k-1)$ of consecutive
segments represents one partial derivative $\partial /\partial K_{i_k
i_{k-1}}$, which we graphically depict by the boundary point (also
numbered by $k$) between the two segments. These $n =: d_0$ points
separating consecutive segments are the 0-cells of $\Sigma(p)$.

The partition $p$ has not yet been used; but now, if $l$ numbers
taken from the set $\{ 1, \ldots, n \}$ constitute one block of $p$,
we draw $l-1$ arcs across $D$ to connect the members of that block
(or rather, the numbered $0$-cells assigned to them) with one
another. We take each such arc to be another $1$-cell of $\Sigma(p)$.
Note that the total number of $1$-cells of $\Sigma(p)$ is
\begin{equation}
    d_1(p) = n + \sum_{l > 1} (l-1) \nu_l(p) \;.
\end{equation}

\begin{figure}
  \begin{center}
    \epsfig{file=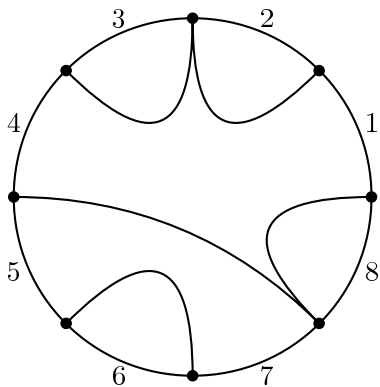,height=5cm}
  \end{center}
    \caption{Example of a non-crossing partition $p = \{ 1, 5, 8\}
    \cup \{ 2, 3, 4 \} \cup \{ 6, 7 \}$ for $n = 8$. 0-cells belonging
    to the same block of the partition are connected by lines. These
    lines divide the disk into sectors (or 2-cells).} \label{fig:bild1}
\end{figure}

Because the partition $p$ is non-crossing, the lines of the $1$-cells
of $\Sigma(p)$ divide the area of the disk $D$ into sectors. We take
these sectors to be the $2$-cells of the complex $\Sigma(p)$, and
denote their total number by $d_2(p)\,$. Figure \ref{fig:bild1} shows
the result of this construction for the example given above.
[Moreover, it is clear that any choice of orientation for the $1$-
and $2$-cells of $\Sigma(p)$ turns $\Sigma(p)$ into a differential
complex $(\Sigma(p),\partial)$ with boundary operator $\partial$. It
follows that $(\Sigma(p),\partial) $ has an Euler characteristic,
which can be computed as the alternating sum $d_0 - d_1(p) +
d_2(p)$.]

We are now in a position to evaluate the contribution to
(\ref{eq:mom-f-cum}) from our fixed partition $p \in \mathrm{NC}(n)$.
According to the definition (\ref{eq:nth-cum}), each block of length
$l$ of derivatives $\partial / \partial K_{i_k i_{k-1}}$ encodes one
cumulant tensor of degree $l\,$. By the identity (\ref{eq:cum-invs})
and the hypothesis (\ref{hypothesis}) the large-$N$ leading
contribution of each such block is a factor of $N^{-l+1} c_l$ times
the sum over irreducible permutations $\sum_{\pi \in [\mathrm{irr}]}
\prod \delta_{i_k , j_{\pi(k)}}$. Note that the total power of $N$
from the product of these factors is
\begin{equation}
    N^{-\sum (l-1) \nu_l(p)} = N^{d_0 - d_1(p)} \;.
\end{equation}

We still need to do the sum over indices $i_1, i_2, \ldots, i_n$ for
the product of Kronecker-delta symbols. Recall that by
(\ref{eq:cum-invs}) the set of lower indices of a nonzero component
of the cumulant tensor must be the same as its set of upper indices.
This constraint forces some of the indices to be equal. [For our
example (\ref{eq:example}) we have $i_1 = i_4$ and $i_5 = i_7$.]
There is a graphical meaning for these constraints: $i_a = i_b$ if
the two 1-cells numbered by $a, b \in \{1, \ldots, n\}$ lie in the
boundary of the same 2-cell of our complex $\Sigma(p)$.

Next we observe that for each block or cumulant tensor there exists
just one large-$N$ optimal permutation $\pi \in \mathrm{S}_l$ in the
sum of (\ref{eq:cum-invs}) -- this is the shift permutation, or
translation by one unit; it is optimal because it produces no further
constraints and hence yields the maximal power of $N$ from index
sums.

We finally calculate the index sum: each of the $d_2(p)$ 2-cells of
$\Sigma(p)$ amounts to one free index giving a factor of $N$; there
are no extra combinatorial factors, as the optimal permutation is
unique for each block of $p$; hence the index sum equals $N^{d_2(p)
}$. It follows that the total contribution to (\ref{eq:mom-f-cum})
from $p \in \mathrm{NC}(n)$ is $\prod c_l^{\nu_l(p)}$ multiplied by
\begin{equation}
    N^{d_0 - d_1(p) + d_2(p)} = N^{\chi(\Sigma(p))}  \;.
\end{equation}
Now the Euler characteristic of a disk $D$ is $\chi(D) = 1$.
Therefore, since our complex $\Sigma(p)$ shares with $D$ its
simplicial homology by construction, we have the relation
\begin{equation}
    \chi(\Sigma(p)) = d_0 - d_1(p) + d_2(p) = 1 \;.
\end{equation}
Thus the total power is $N^1$, which is canceled by the normalization
factor $N^{-1}$ in (\ref{eq:mom-f-cum}).

In summary, the result of summing all contributions from non-crossing
partitions is
\begin{equation}\label{eq:speicher-c}
    m_{n,\,N} = \sum_{p \in \mathrm{NC}(n)}\;
    \prod_{l=1}^n c_l^{\nu_l(p)} + \ldots
\end{equation}
where the dots signify corrections from partitions $p \notin
\mathrm{NC}(n)$ and from the multi-trace terms indicated in
(\ref{multi-trace}).

\subsubsection{Correction terms.}

\begin{figure}
  \begin{center}
    \epsfig{file=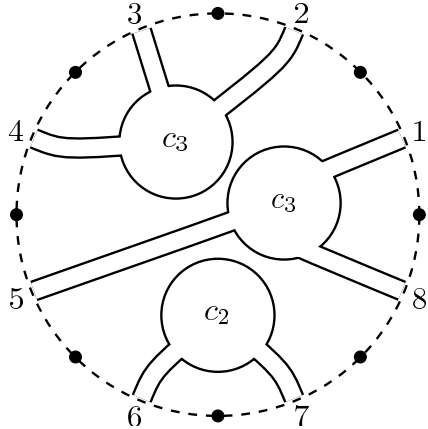,height=5cm}
  \end{center}
    \caption{Another drawing of the non-crossing partition of Figure
    \ref{fig:bild1}. Blocks of size $l$ are now drawn as `blobs' with
    $l$ arms. Each such block represents a free cumulant $c_l\,$.}
    \label{fig:bild2}
\end{figure}

The corrections to (\ref{eq:speicher-c}) become negligible in the
limit $N \to \infty$, and we now briefly discuss why. For that, we
slightly expand our graphical representation. Let still $p \in
\mathrm{NC}(n)$ and consider once more the associated 2-complex
$\Sigma(p)$. Now, however, replace the arcs for each block of $p$ of
length $l$ by a `blob' --- the sum of all perturbation theory graphs
for the cumulant tensor of degree $l$ --- with $l$ external lines
connecting it to the $l$ members of that block. The resulting graph
is planar [see Figure \ref{fig:bild2} for how it looks in the case of
our example (\ref{eq:example})]. It is planar because the partition
$p$ was taken to be non-crossing and the blobs were chosen to be full
disks, as opposed to disks with one or several holes in them.

Now, inserting any correction of the multi-trace form
(\ref{multi-trace}) amounts to changing the topology of a blob from
disk to annulus or higher genus. Our graph then is no longer planar.
The same goes for the replacement of $p \in \mathrm{NC}(n)$ by $p
\notin \mathrm{NC} (n)$: the resulting graph cannot be planar, as
some lines emanating from the blobs must cross. As should be
plausible by now, this loss of planarity results in the loss of at
least one factor of $N$. We therefore claim that all corrections
vanish in the large-$N$ limit:
\begin{equation}
    m_n = \lim_{N\to\infty} m_{n,\,N} = \sum_{p \in \mathrm{NC}(n)}
    \;\prod_{l=1}^n c_l^{\nu_l(p)} \;.
\end{equation}
This is Speicher's formula (\ref{eq:speicher}), provided that we
identify the coefficients $c_n$ -- unknowns for us up to now -- with
the free cumulants of free probability theory.

In summary, assuming that (\ref{eq:expect}) holds, we have argued
that the coefficients $c_n$ must be the free cumulants. This
concludes our perturbation theory argument in support of
(\ref{eq:expect}). Compared to the (non-perturbative) reasoning of
Sections \ref{sect:Dyson} and \ref{sect:FF}, this argument has the
advantage that it applies to the general mixed case where
\begin{equation}
    K = \sum_{a=1}^p \varphi_a \otimes \widetilde{\varphi}_a +
    \sum_{b=1}^q \psi_b \otimes \widetilde{\psi}_b \;.
\end{equation}

\smallskip\noindent\textbf{Note added.}
The large-$N$ hypothesis (\ref{hypothesis}) with $c_n$ equal to the
$n^\mathrm{th}$ free cumulant, is essentially equivalent to Theorem
2.6 of Collins, Mingo, \'Sniady, and Speicher \cite{CMSS}. (We thank
a referee for alerting us to that result.)

\section{Application to Disordered Scattering}
\label{sect:scattering}

As explained in Section \ref{sect:motiv}, the arrival of the
superbosonization method made it possible, in principle, for the
existing treatment of Gaussian ensembles by supersymmetry techniques
to be extended to non-Gaussian ensembles. What was missing up to now
was a good understanding of the large-$N$ behavior of the
characteristic function $\Omega(NK)$. Having developed such an
understanding in the present paper, we are now in a position to go
ahead with the investigation of non-Gaussian ensembles and, in
particular, of questions of universality. A natural candidate for a
first application of the new formalism would be the well-known
universality hypothesis for the spectral correlations of
$\mathrm{U}_N$-invariant ensembles of $N \times N$ Hermitian random
matrices in the large-$N$ limit. This hypothesis, however, has
already been discussed extensively in the literature and strong
results have been obtained by other methods. We therefore refrain
from pursuing this issue here and, instead, turn to a problem which
so far has been inaccessible: the question of universality in models
of stochastic scattering.

To be concrete, we consider a standard scattering problem with $M$
scattering channels coupled to an $N$-dimensional `internal' space
$\mathbb{C}^N$. We will use what is some\-times called the Heidelberg
approximation to the scattering matrix:
\begin{equation}\label{eq:S-matrix}
    S(E) = \mathrm{Id}_M - 2 \mathrm{i} W^\dagger
    \frac{1}{E -H + \mathrm{i}W W^\dagger}\, W \;.
\end{equation}
Here $H$ is the random matrix Hamiltonian acting in the internal
space $\mathbb{C}^N$, the scalar parameter $E$ is the scattering
energy, and $W \in \mathrm{Hom}(\mathbb{C}^M, \mathbb{C}^N)$ is a
coupling operator with Hermitian adjoint $W^\dagger \in \mathrm{Hom}
(\mathbb{C}^N , \mathbb{C}^M)$. Our interest is in the limit $N \to
\infty$ with the number $M$ of scattering channels kept fixed.

To be specific, we consider the correlation function of two elements
of the scattering matrix (correlation functions of higher order can
be treated in exactly the same way):
\begin{equation}\label{eq:def-CFn}
    C_{ab,\,cd}(E_1,E_2) = \left\langle (S_{ab}(E_1) - \delta_{ab})
    (\overline{S_{cd}(E_2)} - \delta_{cd}) \right\rangle \;,
\end{equation}
where the indices label scattering channels, and the angular brackets
denote the expectation value w.r.t.\ to a probability measure $\mu_N$
which we take to be $\mathrm{U}_N$-invariant but non-Gaussian. Since
we eventually want to utilize the result (\ref{eq:result-MZ}), we
require $\mu_N$ to be of the form (\ref{eq:def-mu(H)}) with analytic
and uniformly convex $V$.

Let us note that the correlation function (\ref{eq:def-CFn}) cannot
be expressed solely in terms of the eigenvalues of $H$ or
$H_\mathrm{eff} := H - \mathrm{i} W W^\dagger$. One does have an
alternative expression
\begin{displaymath}
    S(E) = \frac{\mathrm{Id}_M - \mathrm{i} K(E)}
    {\mathrm{Id}_M + \mathrm{i} K(E)} \;, \qquad
    K(E) = W^\dagger (E-H)^{-1} W \;,
\end{displaymath}
in terms of the Wigner-reaction matrix $K(E)$, but there exists no
transparent relation between the distribution of eigenvalues of
$K(E)$ and that of $H$. Therefore, one does not know how to use
orthogonal polynomial techniques for the computation of
(\ref{eq:def-CFn}).

Our method of choice is the supersymmetry method as reviewed in
Section \ref{sect:motiv}. To launch it, we express the elements of
the $S$-matrix as derivatives of a determinant:
\begin{equation}
    S_{ab}(E) - \delta_{ab} = -2 \frac{\partial}{\partial X_{ba}}
    \ln \mathrm{Det}(E - H + \mathrm{i} W X W^\dagger) \Big\vert_{X
    = \mathrm{Id}_M}\;.
\end{equation}
For the complex conjugates $\overline{S_{cd}(E_2)} - \delta_{cd}$ in
(\ref{eq:def-CFn}) we do the same and thus obtain the following
expression for the correlation function:
\begin{eqnarray}
    C_{ab,cd}(E_1,\,E_2) = \frac{- 4\,\partial^2}{\partial X_{ba}
    \partial Y_{dc}} Z(X,Y) \Big\vert_{X = Y = \mathrm{Id}_M} ,\cr
    \hspace{-1cm} Z(X,Y) = \int \frac{\mathrm{Det}(E_1 - H +
    \mathrm{i}W X W^\dagger) \, \mathrm{Det}(E_2 - H - \mathrm{i}
    W W^\dagger)}{\mathrm{Det}(E_1 - H + \mathrm{i}WW^\dagger) \,
    \mathrm{Det}(E_2 - H - \mathrm{i}W Y W^\dagger)}\, d\mu_N(H)\;.
\end{eqnarray}
While the integrand on the right-hand side differs from that of
(\ref{eq:int-formula}) by the presence of the $W W^\dagger$ terms, it
is easy enough to incorporate these into the formalism. We get
\begin{eqnarray}
    Z(X,Y) = \int \Omega(K_1 + K_2) \, \mathrm{e}^{- \mathrm{Tr}\,
    (E_1 K_1 + E_2 K_2)}\cr \times \mathrm{e}^{-\mathrm{i} \langle
    \widetilde{\varphi}_1 , W W^\dagger\varphi_1 \rangle + \mathrm{i} \langle
    \widetilde{\psi}_1,\, W X W^\dagger \psi_1 \rangle + \mathrm{i} \langle
    \widetilde{\varphi}_2 ,\, W Y W^\dagger \varphi_2 \rangle - \mathrm{i}
    \langle \widetilde{\psi}_2 , \, W W^\dagger \psi_2 \rangle}\;,
\end{eqnarray}
where $K_j = \varphi_j \otimes \widetilde{\varphi}_j + \psi_j \otimes
\widetilde{\psi}_j$ ($j = 1, 2$). The integral sign stands for the
Berezin integral over the anti-commuting variables of $\psi_j,
\widetilde{\psi}_j$ as well as the ordinary integral (with Lebesgue
measure) over $\varphi_j, \widetilde{\varphi}_j\,$. The integration
domain for the latter consists of two copies ($j = 1, 2$) of
$(\mathbb{C}^N)^\ast \times \mathbb{C}^N$ restricted to the real
subspace $\mathbb{C}^N$ which is given by $\widetilde{\varphi}_1 = -
\mathrm{i} \varphi_1^\dagger$ and $\widetilde{\varphi}_2 = +
\mathrm{i} \varphi_2^\dagger\,$, respectively.

We now take the derivatives $\partial /\partial X_{ba}$ and $\partial
/ \partial Y_{dc}$ at $X = Y = \mathrm{Id}_M\,$. We also scale $K_j
\to N K_j\,$. The expression for the correlation function then
becomes
\begin{eqnarray}
    C_{ab,\,cd}(E_1,E_2) = \int \Omega(N K_1 + N K_2)\,\mathrm{e}^{
    - N\, \mathrm{Tr}\, (E_1 K_1 + E_2 K_2)} \label{eq:6.4} \\
    \times 4N^2 \mathrm{e}^{- N\, \mathrm{Tr}\;(\mathrm{i}K_1 -
    \mathrm{i}K_2) W W^\dagger} \langle \widetilde{\psi}_1,\, W_b
    \rangle \langle W_a^\dagger , \psi_1 \rangle \langle
    \widetilde{\varphi}_2 , W_d \rangle \langle W_c^\dagger, \varphi_2
    \rangle .
\end{eqnarray}

At this stage, we would like to employ the superbosonization formula
(\ref{eq:supbos}), using the large-$N$ information about $\Omega(NK)$
from Sections \ref{sect:Dyson}-\ref{sect:combinatorics}. However,
this is not immediately possible, as the second line of
(\ref{eq:6.4}) is not expressed by $\mathrm{GL}_N$-invariants owing
to the presence of $W$ and $W^\dagger$. We therefore use a trick.

\subsection{Averaging trick}

The trick is to \emph{enforce} invariance by making a substitution of
integration variables (which has unit Jacobian),
\begin{equation}
    \varphi_j \to g \varphi_j \;, \quad \psi_j \to g \psi_j \;, \quad
    \widetilde{\varphi}_j \to \widetilde{\varphi}_j \, g^{-1} \;, \quad
    \widetilde{\psi}_j \to \widetilde{\psi}_j \, g^{-1} \;.
\end{equation}
and averaging over $g \in \mathrm{U}_N$ with Haar measure to define
the auxiliary quantity
\begin{eqnarray}
    D_{ab,\,cd}^{(N)} := 4N^2 \int_{\mathrm{U}_N}
    \mathrm{e}^{-N\,\mathrm{Tr}\; g\, (\mathrm{i}K_1 - \mathrm{i}K_2)
    \, g^{-1} W W^\dagger} \cr \times\langle \widetilde{\psi}_1\,
    g^{-1},\, W_b \rangle \langle W_a^\dagger , g \psi_1 \rangle
    \langle \widetilde{\varphi}_2 \, g^{-1} , W_d \rangle \langle
    W_c^\dagger , g \varphi_2 \rangle \, dg \label{eq:mrz-xyz}\;.
\end{eqnarray}

Our next step is to compute the unitary matrix integral $D_{ab,\,cd}
^{(N)}$ in the large-$N$ limit. For this we use the following result:
if $A, B \in \mathrm{End}(\mathbb{C}^N)$ are operators whose rank is
kept fixed (i.e., finite) in the limit $N \to \infty$, then
\begin{equation}\label{eq:limit-law}
    \int_{\mathrm{U}_N}\mathrm{e}^{-N\,\mathrm{Tr}\,(A\,g\,B\,g^{-1})}
    dg \simeq \mathrm{Det}^{-1}(\mathrm{Id}_N \otimes\mathrm{Id}_N +
    A\otimes B)\;,
\end{equation}
where `$\simeq$' means equality in the large-$N$ limit, with the
right-hand side tending to the determinant of the Fredholm operator
$\mathrm{End}(\mathbb{C}^\infty) \ni X \mapsto X + A X B\,$. While
this formula follows as a corollary of the relation between the
integral $\int \mathrm{e}^{ -N\,\mathrm{Tr}\, (A\,g\, B\,g^{-1})}dg$
and the $R$-transform \cite{pzj-PRE}, it can be obtained more
directly by the observation that the matrix entries of $g$ and
$g^{-1} = g^\dagger$ (under the fixed rank condition on both $A$ and
$B$) become independent complex Gaussian random variables of variance
$N^{-1}$ in the large-$N$ limit.

We now apply (\ref{eq:limit-law}) to the present situation with $A =
W W^\dagger$ and $B = \mathrm{i}(K_1 - K_2):$
\begin{eqnarray}
    \int_{\mathrm{U}_N} \mathrm{e}^{-N\,\mathrm{Tr}\;(W W^\dagger g\,
    (\mathrm{i}K_1 - \mathrm{i}K_2)\,g^{-1})} dg\cr \simeq
    \mathrm{Det}^{-1}(\mathrm{Id}_N \otimes \mathrm{Id}_N + WW^\dagger
    \otimes \mathrm{i}(K_1-K_2))\;.
\end{eqnarray}
Then we switch from the determinant on $\mathbb{C}^N \otimes
\mathbb{C}^N$ to a (super-)determinant on $\mathbb{C}^M \otimes
\mathbb{C}^{2|2}$, using that $W^\dagger$ and $W$ exchange
$\mathbb{C}^N$ with $\mathbb{C}^M$, while the quadruples
$\widetilde{\varphi}_1, \widetilde{\varphi}_2, \widetilde{\psi}_1,
\widetilde{\psi}_2$ and $\varphi_1, \varphi_2, \psi_1, \psi_2$
exchange $\mathbb{C}^N$ with the $\mathbb{Z}_2$-graded vector space
$\mathbb{C}^{2|2}:$
\begin{eqnarray}
    \lim_{N\to\infty} \int_{\mathrm{U}_N} \mathrm{e}^{-N\,\mathrm{Tr}
    \;(W W^\dagger g\,(\mathrm{i}K_1 - \mathrm{i}K_2)\,g^{-1})} dg\cr
    = \mathrm{SDet}^{-1}(\mathrm{Id}_M \otimes \mathrm{Id}_{2|2} +
    W^\dagger W \otimes \mathrm{i}s Q) \;.
\end{eqnarray}
Here $s = \mathrm{diag}(1,-1,1,-1)$ and $Q$ denotes the supermatrix
of $\mathrm{GL}_N$-invariants
\begin{equation}\label{eq:def-Q}
    Q = \left( \begin{array}{llll}
     \langle \widetilde{\varphi}_1 , \varphi_1 \rangle
    &\langle \widetilde{\varphi}_1 , \varphi_2 \rangle
    &\langle \widetilde{\varphi}_1 , \psi_1 \rangle
    &\langle \widetilde{\varphi}_1 , \psi_2 \rangle \cr
     \langle \widetilde{\varphi}_2 , \varphi_1 \rangle
    &\langle \widetilde{\varphi}_2 , \varphi_2 \rangle
    &\langle \widetilde{\varphi}_2 , \psi_1 \rangle
    &\langle \widetilde{\varphi}_2 , \psi_2 \rangle \cr
     \langle \widetilde{\psi}_1 , \varphi_1 \rangle
    &\langle \widetilde{\psi}_1 , \varphi_2 \rangle
    &\langle \widetilde{\psi}_1 , \psi_1 \rangle
    &\langle \widetilde{\psi}_1 , \psi_2 \rangle \cr
     \langle \widetilde{\psi}_2 , \varphi_1 \rangle
    &\langle \widetilde{\psi}_2 , \varphi_2 \rangle
    &\langle \widetilde{\psi}_2 , \psi_1 \rangle
    &\langle \widetilde{\psi}_2 , \psi_2 \rangle
    \end{array} \right) \;.
\end{equation}

Next, to account for the post-exponential factors in the integral
(\ref{eq:mrz-xyz}) we introduce
\begin{equation}
    f(Q) = (W^\dagger W \otimes Q)\, \big( \mathrm{Id}_N \otimes
    \mathrm{Id}_{2|2} + W^\dagger W\otimes \mathrm{i}sQ \big)^{-1}.
\end{equation}
By a slight extension of (\ref{eq:limit-law}) to include these
factors, we have
\begin{eqnarray}
    \lim_{N\to\infty} D_{ab,\,cd}^{(N)} = \mathrm{SDet}^{-1}
    (\mathrm{Id}_M \otimes \mathrm{Id}_{2|2} + W^\dagger W
    \otimes \mathrm{i}s Q) \cr \times 4 \left( f(Q)_{ab;\,
    \widetilde{\psi}_1,\psi_1}\,f(Q)_{cd;\widetilde{\varphi}_2,
    \varphi_2}+ f(Q)_{cb\,;\widetilde{\psi}_1,\varphi_2}\,
    f(Q)_{ad;\widetilde{\varphi}_2,\psi_1}\right)\;,
\end{eqnarray}
where the first index pair indicates the position of the matrix entry
in $\mathrm{End}(\mathbb{C}^M)$, and the second pair indicates the
one in $\mathrm{End}(\mathbb{C}^{2|2})$. In the following we write
\begin{equation}
    \lim_{N\to\infty} D_{ab,\,cd}^{(N)} =: F_{ab,\,cd}(Q) \;.
\end{equation}

\subsection{Superbosonization}\label{sect:6.2}

Since the averaging trick has converted all dependence on
$\varphi,\psi$ of the integrand in (\ref{eq:6.4}) into an implicit
dependence through the supermatrix $Q$ of (\ref{eq:def-Q}), we can
now apply the superbosonization formula (\ref{eq:supbos}). In the
superbosonization step one forgets the expression (\ref{eq:def-Q})
for the supermatrix $Q$ and treats the matrix entries of $Q$ as the
new integration variables; at the same time, the integrand is
`lifted' to a function of $Q$.

To bring the result of superbosonization into a form suitable for the
large-$N$ saddle analysis of Section \ref{sect:6.3}, we write
$\mathrm{SDet}^N(Q) = \mathrm{e}^{N \ln\,\mathrm{SDet}\, Q} =
\mathrm{e}^{N\,\mathrm{STr}\,\ln Q}$. With the same motivation, we
write the lift $\widehat{\Omega}(NQ)$ of the characteristic function
$\Omega(NK)$ of (\ref{eq:expect}) as
\begin{equation}
    \widehat{\Omega}(NQ) = \mathrm{e}^{N\,\mathrm{STr}\;\Phi(Q)+O(N^{-1})}
    \;,\quad \Phi(Q) = \sum\nolimits_{n=1}^\infty \frac{c_n}{n}\;Q^n\;.
\end{equation}
By sending $N$ to infinity, we then obtain the exact result
\begin{eqnarray}
    \lim_{N\to \infty} C_{ab,\,cd}(z+\varepsilon/N,z-\varepsilon/N)\cr
    = \lim_{N\to\infty} \int DQ \; \mathrm{e}^{N\,\mathrm{STr}\,(\ln Q
    + \Phi(Q) - z\,Q)} \mathrm{e}^{-\varepsilon\, \mathrm{STr}\,(sQ)}
    F_{ab,\,cd}(Q).
\end{eqnarray}
Here we used the fact (cf.\ \cite{LSZ-D2}) that the normalization
constant in (\ref{eq:supbos}) has the value $c_{p,\,q} = 1$ for the
present case of $p = q = 2$. The explicit expression for the
invariant Berezin integration form $DQ$ defined in (\ref{eq:DQ})
becomes
\begin{equation}
    DQ = (2\pi)^{-4} \prod_{i,j=1}^2 dQ_{\widetilde{\varphi}_i,
    \varphi_j}\, dQ_{\widetilde{\psi}_i,\psi_j} \frac{\partial^2}
    {\partial Q_{\widetilde{\varphi}_i,\psi_j}\,
    \partial Q_{\widetilde{\psi}_i,\varphi_j}}\;.
\end{equation}
We also observe that by (\ref{eq:Hps}) the $Q_{\widetilde{\varphi}
\varphi}$-integral ranges over $(\mathrm{i} s Q)_{\widetilde{\varphi}
\varphi}^\dagger = (\mathrm{i} s Q)_{\widetilde{\varphi}\varphi} >
0\,$, while the $Q_{\widetilde{\psi}\psi}$-integral ranges over
$\mathrm{U}_2\,$.

One may ask why we keep $\mathrm{e}^{-\varepsilon\, \mathrm{STr} \,
(sQ)} F_{ab,\,cd}(Q)$ but neglect $O(N^0)$ correction terms in $\ln
\widehat{\Omega}(NQ)$ which appear to be of the same order. This
question will be answered at the end of the next subsection.

\subsection{Saddle approximation}\label{sect:6.3}

In the limit $N \to \infty$ the dominant factor in the integral is
$\mathrm{e}^{N\,\mathrm{STr}\,(\ln Q + \Phi(Q) - zQ)}$. Taking the
first variation of the exponent gives the saddle-point equation
\begin{equation}\label{eq:speqn}
    Q^{-1} + R(Q) = z \cdot \mathrm{Id}_{2|2} \;, \qquad
    R(Q) = \sum_{n=1}^\infty c_n\, Q^{n-1} \;.
\end{equation}
This is Voiculescu's equation (\ref{eq:Voiculescu}), except that the
role of the complex variable $k$ is now taken by the supermatrix $Q$.

Let $z \in \mathbb{R}$ be inside the support of the large-$N$
spectral measure of the random matrix $H$. Then we know that the
equation $q^{-1} + R(q) = z$ for a scalar variable $q \in \mathbb{C}$
has two solutions, $q = g_+(z)$ and $q = g_-(z)$, with
\begin{equation}
    \mathfrak{Re}\, g_+(z) = \mathfrak{Re}\, g_-(z) \;, \quad
    \mathfrak{Im}\, g_+(z) = - \mathfrak{Im}\, g_-(z) < 0 \;.
\end{equation}

The following analysis is standard \cite{VWZ} and we therefore give
only a sketch. We first look for diagonal matrices $Q$ that solve the
saddle-point equation (\ref{eq:speqn}). The condition $\mathrm{i} s
Q_{\widetilde{\varphi}\varphi} > 0$ selects
\begin{equation}
    Q_{\widetilde{\varphi}\varphi} = \mathrm{diag}\,(g_+(z),g_-(z)) \;.
\end{equation}
(This is literally true if $\mathfrak{Re}\, g_{\pm}(z) = 0$. When the
real part is nonzero, a contour deformation is necessary in order to
reach this saddle point.)

In the fermion-fermion sector (i.e., $Q_{\widetilde{\psi}\psi}$)
there are in principle four possible choices of diagonal saddle for
$Q_{\widetilde{\psi}\psi}$. Two of these, $(++)$ and $(--)$, do not
contribute for $N \to \infty$ as they come with suppression factors
$1/N$ due to fermionic Goldstone modes from breaking of
supersymmetry. The remaining two choices, $(+-)$ and $(-+)$, are
equivalent in the sense that they turn out to lie on the same orbit
of the symmetry group. Thus we select the diagonal saddle
\begin{equation}
    \hspace{-1cm} Q_0 = \mathrm{diag}\,(g_+(z),g_-(z),g_+(z),g_-(z))
    = \mathfrak{Re}\, g_+(z) \mathrm{Id}_{2|2} + \mathrm{i}
    \mathfrak{Im}\, g_+(z) \, s \;.
\end{equation}

The dominant part of the integrand is invariant under conjugation $Q
\to T Q T^{-1}$ by elements $T$ of the Lie supergroup $\mathrm{U}_{1,
1|2}$ \cite{berezin}. The orbit generated by the action on $Q_0$ of
this symmetry group is a supermanifold of saddle points
\begin{equation}
    Q = T Q_0(z) T^{-1} = \mathfrak{Re}\, g_+(z) \mathrm{Id}_{2|2}
    + \mathrm{i} \mathfrak{Im} \,g_+(z) \, T s T^{-1} \;.
\end{equation}
Evaluating the dominant part of the integrand along this manifold we
simply get unity:
\begin{equation}
    \mathrm{e}^{N\,\mathrm{STr}\,(\ln Q + \Phi(Q) - z\,Q)}
    \big\vert_{Q = T Q_0(z) T^{-1}} = \mathrm{e}^0 = 1 \;.
\end{equation}
Moreover, the integration over the Gaussian fluctuations normal to
the saddle-point manifold also gives unity by supersymmetry. Hence we
have
\begin{eqnarray}
    \lim_{N\to \infty} C_{ab,\,cd}(z+\varepsilon/N,z-\varepsilon/N)\cr =
    \int DT \; \mathrm{e}^{-\varepsilon\, \mathrm{STr}\,(s T Q_0(z) T^{-1})}
    F_{ab,\,cd}(T Q_0(z) T^{-1}) \;,
\end{eqnarray}
where $DT$ denotes the $\mathrm{U}_{1,1|2}$-invariant Berezin
integration form for the saddle-point supermanifold. This is the very
same result which is obtained for the case of the Gaussian Unitary
Ensemble in the limit $N \to \infty\,$. In this sense the result is
universal.

Let us summarize what are the agents of the mechanism leading to
universality. First of all, the dominant factor $\mathrm{e}^{
N\,\mathrm{STr}\,(\ln Q + \Phi(Q) - z\,Q)}$ of our integral is
invariant under a Lie supergroup $\mathrm{U}_{1,1|2}\,$. This
symmetry group is determined by the type of correlation function
under consideration and does not depend on the details of the
probability measure $\mu_N\,$. All saddle-point supermanifolds then
are orbits of $\mathrm{U}_{1,1|2}\,$.

Second, by the principle of maximal supersymmetry the large-$N$ limit
always selects the same type of orbit, $Q = T Q_0(z) T^{-1}$, as long
as $z$ lies in the bulk of the spectrum. [At the edges of the
spectrum, the orbit degenerates by the vanishing of $\Delta := \mp
\mathfrak{Im}\, g_{\pm}(z)$.] The specific details of the probability
measure $\mu_N$ enter only via the scale factor $\Delta\,$.

Third, the correlation function is obtained by restricting the
(non-invariant part of the) integrand to the orbit $Q = T Q_0(z)
T^{-1}$ and integrating along it. The scale factor $\Delta$ is needed
in order to express the energy dependence of the correlation function
in the proper units given by the mean level spacing. At the same
time, the `scattering observable' $F_{ab,\,cd}(T Q_0(z) T^{-1})$ is
expressed in terms of the average $S$-matrix and physical quantities
called `sticking probabilities' or transmission factors \cite{VWZ}.
When this is done, the correlation function assumes its universal
form.

Let us finally give the reason why it was legitimate to neglect the
$O(N^0)$ corrections to $\ln \widehat{\Omega}(NQ)$. Any such
correction is $\mathrm{U}_{1,1|2}$-invariant and, in fact, vanishes
along the dominant saddle-point supermanifold. Its only effect is to
cause a slight perturbation of the scale of this supermanifold and
correct the density of states by a term of order $1/N$. This effect
is negligible in the large-$N$ limit.

\section{Summary and outlook}\label{sect:last}

In this paper, we employed a variety of techniques to study the
characteristic function $\Omega(NK)$ and its lift
$\widehat{\Omega}(NQ)$, which are key to a recent variant of the
supersymmetry method, i.e., the Wegner-Efetov technique of
integration over commuting and anti-commuting variables. What we
found is that the large-$N$ asymptotics of $\widehat{\Omega}(NQ)$ for
any unitary ensemble `close to Gaussian' is determined by the
$R$-transform known from free probability theory. (More precisely, we
made the assumption that the confining potential for the random
matrix eigenvalues is uniformly convex and analytic, in which case
the $R$-transform is an entire function.) The task of computing
correlation functions then reduces to a discussion centered around
supermanifolds which are given as solution spaces of Voiculescu's
equation (\ref{eq:speqn-V}) extended to the case of a supermatrix
$Q\,$.

This insight opens the door to numerous applications which in the
past had been beyond the reach of the supersymmetry method. In the
present paper, we have given a first application to stochastic
scattering, demonstrating the universality of $S$-matrix correlations
for the case of unitary ensembles close to Gaussian. Future
applications of the method will be aimed at more demanding situations
with two or more cuts (i.e., with a density of states supported on
several disjoint intervals) and double scaling limits at critical
points. To make progress with such far-from-Gaussian problems, we
first have to learn how to deal with singularities that develop in
the $R$-transform.

We stress that although our paper deals exclusively with unitary
ensembles, the methods used are robust and do extend to ensembles of
different symmetry type.

Let us finish with a quick glance at a new and exciting development.
In a long series of papers by Erd\"os, Ramirez, Schlein, Yau, and
Tao, Vu \cite{series1,series2,TV,gang6,final}, sine-kernel (or GUE)
universality of spectral correlations has recently been established
for the case of Hermitian \emph{Wigner} matrices, i.e., random
matrices with statistically independent entries. Using these results
as input to the present formalism, we may now address the wider class
of random matrices given as the sum of a Wigner matrix and a unitary
ensemble. Indeed, the characteristic function of such a random matrix
is a product $\Omega =\Omega_{\rm Wigner}\times\Omega_{\rm unitary}$.
In the present paper we developed a large-$N$ theory of the second
factor, while the results of Erd\"os et al.\ give control of the
first factor; more precisely, $\widehat{\Omega}_{\mathrm{Wigner}}
(NQ)$ approaches $\widehat{\Omega}_{\mathrm{GUE}} (NQ)$ with a rate
of convergence which is sufficiently fast in order for sine-kernel
universality to emerge. Moreover, because the sum of a Wigner matrix
and a unitary ensemble is a sum of free random variables, the
$R$-transform of the sum is the sum of the individual $R$-transforms.
This offers a good prospect of obtaining analytical control of the
more general situation.

\medskip\noindent\textbf{Acknowledgment}. M.R.Z.\ acknowledges a
helpful discussion with A.\ Guionnet. This work was financially
supported by the Deutsche Forschungsgemeinschaft (SFB/TR 12).


\section*{References}
\begin{thebibliography}{11}
%
\bibitem{BDS} Baik J, Deift P and Strahov E 2003 Products
    and ratios of characteristic polynomials of random Hermitian
    matrices {\it J.\ Math.\ Phys.} {\bf 44} 3657
%
  \bibitem{berezin} Berezin FA 1987 {\it Introduction to
    Superanalysis} (Dordrecht: Reidel)
%
\bibitem{BGV} Berline M, Getzler E and Vergne M 1992 {\it Heat
    kernels and Dirac operators} (Berlin: Springer)
%
\bibitem{large-N} Brezin E, Itzykson C, Parisi G and Zuber JB
    1978 Planar diagrams {\it Commun. Math. Phys.} {\bf 69} 35
%
\bibitem{bump} Bump D 2004 {\it Lie groups} Graduate Texts in
    Mathematics, vol.\ 225 (Berlin: Springer)
%
\bibitem{collins} Collins B 2003 Moments and cumulants of
    polynomial random variables on unitary groups, the Itzykson-Zuber
    integral and free probability {\it Int. Math. Res. Not.}
    {\bf 17} 953
%
\bibitem{CMSS} Collins B, Mingo JA, \'Sniady P, and Speicher R
    2007 Second order freeness and asymptotics of random matrices
    III. higher order freeness and free cumulants {\it Documenta
    Math.} {\bf 12} 1
%
\bibitem{deift} Deift P 2000 {\it Orthogonal polynomials and random
    matrices: a Riemann-Hilbert approach} Courant Lecture Notes
    (New York: AMS)
%
\bibitem{DKMVZ} Deift P, Kriecherbauer T, McLaughlin KTR,
    Venakides S and Zhou X 1999 Uniform asymptotics for polynomials
    orthogonal with respect to varying exponential weights and
    applications to universality questions in random matrix theory
    {\it Commun. Pure Appl. Math.} {\bf 52} 1335
%
\bibitem{efetov} Efetov KB 1983 Supersymmetry and theory of
    disordered metals {\it Adv. Phys.} {\bf 32} 127
%
\bibitem{series1} Erdos L, Ramirez J, Schlein B and Yau H-T
    2009 Universality of sine kernel for Wigner matrices with a
    small Gaussian perturbation {\it Preprint} arXiv:0905.2089
%
\bibitem{series2} Erdos L, Ramirez J, Schlein B and Yau H-T 2009
    Bulk universality for Wigner matrices {\it Preprint} arXiv:0905.4176
%
\bibitem{gang6} Erdos L, Ramirez J, Schlein B, Tao T, Vu V and Yau J-T
    2009 Bulk universality of Wigner hermitian matrices with
    subexponential decay {\it Preprint} arXiv:0906.4400
%
\bibitem{final} Erdos L, Schlein B and Yau H-T 2009
    Universality of random matrices and local relaxation flow
    {\it Preprint} arXiv:0907.5605
%
\bibitem{GM} Guionnet A and Maida M 2005 A Fourier view on the
    $R$-transform and related asymptotics of spherical integrals
    {\it J. Func. Anal.} {\bf 222} 435
%
\bibitem{howeRCIT} Howe R 1989 Remarks on classical invariant
    theory {\it Trans. Amer. Math. Soc.} {\bf 313} 539
%
\bibitem{LSZ-D2} Littelmann P, Sommers H-J and Zirnbauer MR 2008
    Superbosonization of invariant random matrix ensembles
    {\it Commun. Math. Phys.} {\bf 283} 343
%
\bibitem{LSZ-C3} Lueck T, Sommers H-J and Zirnbauer MR 2006
    Energy correlations for a random matrix model of disordered
    bosons {\it J. Math. Phys.} {\bf 47} 103304
%
\bibitem{speicher} Speicher R 1994 Multiplicative functions on
    the lattice of noncrossing partitions and free convolution
    {\it Math. Ann.} {\bf 298} 611
%
\bibitem{szego} Szeg\"o G 1975 {\it Orthogonal polynomials}
    ({\it AMS Colloquium Publications, vol.\ 23})
    4th edn (Providence, RI: American Mathematical Society)
%
\bibitem{TV} Tao T and Vu V 2009 Random matrices: universality
    of local eigenvalue statistics {\it Preprint} arXiv:0906.0510
%
\bibitem{VWZ} Verbaarschot JJM, Weidenm\"uller HA and
    Zirnbauer MR 1985 Grassmann variables in stochastic quantum
    physics: the case of compound-nucleus scattering {\it Phys. Rep.}
    {\bf 129} 367
%
\bibitem{voicul1} Voiculescu D 1986 Addition of certain
    non-commuting random variables {\it J. Func. Anal.} {\bf 66} 323
%
\bibitem{voicul2} Voiculescu D 1991 Limit laws for random
    matrices and free products {\it Invent. math.} {\bf 104} 201
%
\bibitem{VDN} Voiculescu D, Dykema KJ and Nica A 1992 \textit{Free
    random variables} ({\it CRM Monograph Series, vol.~1})
    (Providence, RI: American Mathematical Society)
%
\bibitem{wegner} Wegner F 1979 The mobility edge problem:
    continuous symmetry and a conjecture {\it Z. Phys. B} {\bf 35} 207
%
\bibitem{weyl} Weyl H 1939 \textit{The Classical Groups} (Princeton:
    Princeton University Press)
%
\bibitem{zee} Zee A 1996 Law of addition in random matrix
    theory {\it Nucl. Phys. B} {\bf 474} 726
%
\bibitem{pzj-CMP} Zinn-Justin P 1998 Universality of
    correlation functions of Hermitian random matrices in an external
    field {\it Commun. Math. Phys.} {\bf 104} 631
%
\bibitem{pzj-PRE} Zinn-Justin P 1999 Adding and multiplying
    random matrices: generalization of Voiculescu's formulas
    {\it Phys. Rev. E} {\bf 59} 4884
%
\end{thebibliography}
\end{document}